\documentclass[letter,12pt]{article}

\usepackage{graphicx,amssymb}

\newcommand{\idk}{\int \frac{{\rm d}^3\bk}{(2\pi)^3}}

\newcommand\nn{\nonumber \\}
\newcommand\beq{\begin{equation}}
\newcommand\eeq{\end{equation}}
\newcommand\beqa{\begin{eqnarray}}
\newcommand\eeqa{\end{eqnarray}}

\newcommand{\ds}[1]{#1 \hspace{-0.5em}/}  
\newcommand\bzeta{\mbox{\boldmath$\zeta$}}
\newcommand\bgamma{\mbox{\boldmath$\gamma$}}

\newcommand\bk{{\bf k}}
\newcommand\bq{{\bf q}}
\newcommand\E{\epsilon}

\newcommand\bp{{\bf p}}

\def\sla{\slash{\!\!\!} }
\newcommand{\vp}{\mbox{\boldmath $p$}}

\begin{document}

\title{Magnetism and superconductivity\\
 in quark matter}
\author{Toshitaka Tatsumi$^a$,  Eiji Nakano$^{b}$ and Kanabu Nawa$^a$  
\\
$^a$ {\it\normalsize Department of Physics, Kyoto University, Kyoto 606-8502, Japan}
\\
$^b$  {\it\normalsize Yukawa Institute for Theoretical Physics, Kyoto University, Kyoto
 606-8502, Japan}
}

\maketitle

\begin{abstract}
Magnetic properties of quark matter and its relation to the microscopic 
origin of the magnetic field observed in compact stars are 
studied. Spontaneous spin polarization appears in
high-density region due to the Fock exchange term, which may provide a 
scenario for the behaviors of magnetars.
On the other hand,
quark matter becomes unstable to form spin density wave in the moderate 
 density region, where restoration of chiral symmetry plays an important 
 role. Coexistence of magnetism and color superconductivity is also 
discussed. 

\end{abstract}

\section{Introduction}

QCD has been believed to be the basic theory of strong interaction and 
there are many successful consequences about the properties of hadrons
and their interactions. Recently 
many studies have been devoted to figure out the phase diagram of QCD 
in temperature ($T$) - density ($\rho_B$) plane \cite{alf,lat}. 
At high temperature or high density, quarks confined inside hadrons
should be liberated to form matter consisting of quarks and gluons 
(deconfinement transition). Such basic constituents, especially quarks
exhibit interesting properties there as electrons in condensed matter
through many-body dynamics; one of the interesting possibility is phase
transition as temperature or density changes.   
When we emphasize
the low $T$ and high $\rho_B$ region, the subjects are sometimes called
high-density QCD. 
The main aims in this field should be 
to elucidate the new phases and their properties, and to
extract their symmetry breaking pattern and low-energy excitation modes
there on the basis of QCD. On the other hand, these studies have
phenomenological implications on relativistic heavy-ion collisions and
compact stars like neutron stars or quark stars \cite{tat05}.

Color superconductivity (CSC) should be very popular \cite{alf,bai}. 
Its mechanism is similar to the BCS theory for the 
electron-phonon system \cite{BCS}, 
in which the attractive interaction of electrons is 
provided by phonon exchange and 
causes the Cooper instability near the Fermi surface. 
As for quark matter, 
the quark-quark interaction is mediated by colored gluons, and is 
often approximated by some effective interactions, e.g., 
the one-gluon-exchange (OGE) or the instanton-induced interaction, 
both of which give rise to the attractive quark-quark interaction 
in the color anti-symmetric $\bar 3$ channel. 
Many people believe that it is robust due
to the Cooper instability even for small attractive quark-quark
interaction in color $\bar 3$ channel. 

Here we'd like to address another interesting property of quark matter, 
magnetic properties of quark matter. 
We shall see various types of magnetic ordering may be expected in quark 
matter at finite density or temperature. They arise due to the quark 
particle-hole ($p-h$) 
correlations in the pseudo-scalar or axial-vector channel.

Phenomenologically the concept of magnetism should be directly related
to the origin of strong magnetic field observed 
in compact stars \cite{MAG3}; e.g., it
amounts to $O(10^{12}$G) at the surface of radio pulsars. Recently a new
class of pulsars called magnetars has been discovered with super strong
magnetic field, $B_s\sim 10^{14 - 15}$G, estimated from the $P-\dot{P}$
curve \cite{tho04}. First observations  are indirect evidences for 
the superstrong magnetic field, but 
discoveries of some absorption lines stemming from the cyclotron
frequency of protons have been currently reported \cite{ibr02}; when it
is confirmed that they originate from protons, they give a direct
evidence for the superstrong magnetic field. 

\begin{table}[h]
\caption{Surface magnetic field and the radius of stars by the
 conservation of the magnetic flux. }
\centering
\begin{tabular*}{5cm}{@{\extracolsep{\fill}}lll}
\hline
 ~& $B_S[{\rm G}]$ & $R [{\rm cm}]$\\ \hline
{\rm Sun~(obs.)} & $10^3$ & $10^{10}$\\
{\rm Neutron star} & $10^{11}$ & $10^6$\\ \hline
{\rm Magnetar} & $10^{15}$ & $10^4$
\end{tabular*}
\end{table}

The origin of the
strong magnetic field in compact stars has been a long standing problem 
since the first discovery of a pulsar \cite{MAG3}. 
A simple working hypothesis is the
conservation of the magnetic flux and its squeezing during the evolution
from a main-sequence progenitor star to a compact star; $B\propto R^{-2}$ with
$R$ being the radius. Taking the sun as a typical main-sequence star, we have
$B\sim 10^3$G and $R\sim 10^{10}$cm. If it is squeezed to a typical
radius of usual neutron stars, $R\sim 10$km, the conservation of the
magnetic flux gives $10^{11}$G, which is consistent with the
observations for radio pulsars. However, we find $R\sim 100$m to explain
$B\sim 10^{15}$G observed for magnetars, which may lead to a contradiction
since the Schwatzschild radius is $O(1$km) for the canonical mass of
$O(M_\odot)$, which is much larger than $R$.   

Since dense hadronic matter should be widely developed inside compact
stars, it would be reasonable to inquire a microscopic origin of such
strong magnetic field: ferromagnetism (FM) or spin polarization is one of the
candidates to explain it. Makishima also suggested the hadronic origin
of the magnetic field observed in binary X-ray pulsars or radio pulsars\cite{mak03},
since it looks no field decay in these objects.

When we consider the magnetic-interaction energy by a simple formula,  
$E_{\rm mag}=\mu_iB$ with
the magnetic moment, $\mu_i=e_i/(2m_i)$, we can easily estimate it for
$B=O(10^{15}$G) (see Table 2); it amounts to 
several MeV for electrons, while several keV for nucleons and 10 keV- 1MeV
for quarks dependent on their mass.
\begin{table}[h]
\caption{Magnetic interaction energies $E_{\rm mag}$ for $10^{15}$G 
and the typical energy
 scales $E_{\rm typ}$ in electron, nucleon and quark systems.}
\centering
\begin{tabular*}{10cm}{@{\extracolsep{\fill}}lccc}
\hline
 &electron & proton & quark~~~~\\ \hline
$m_i[{\rm MeV}]$& 0.5 & $10^3$ & 1- 100~~~~\\
$E_{\rm mag}[{\rm MeV}]$~~~ & 5 - 6  &~~~$2.5\times 10^{-3}$~~~& $2.5\times 10^{-2} -
2.5$\\ \hline
$E_{\rm typ}$ & {\rm keV} & {\rm MeV} & {\rm MeV}~~~~
\end{tabular*}
\end{table}
This simple
consideration may imply that strong interaction gives a feasible origin 
for the strong magnetic field, since its typical energy scale is MeV. 
The possibility of
ferromagnetism in nuclear matter has been elaborately studied since the
first discovery of pulsars, 
but negative results have been reported so far \cite{fan01}.  
We consider here its possibility in quark matter as an alternative 
in light of recent development 
of high-density QCD \cite{tat00}.

If FM is realized in quark matter, there should be some 
interplay with CSC;  
we examine a possibility of the coexistence of FM 
and CSC in quark matter, where we shall see an interplay between 
particle-particle and particle-hole correlations. 
As far as we know, 
interplay between the color superconducting phase 
and other phases characterized 
by the non-vanishing mean fields of the spinor bilinears has not been explored 
except for the case of chiral symmetry breaking \cite{CSC3}.  

It would be worth mentioning in this context that  
ferromagnetism (or spin polarization) and superconductivity are 
fundamental concepts in condensed matter physics, and  
their coexistent phase has been discussed and expected for a long time \cite{MagSup1}.
As a recent progress, superconducting phases have been discovered 
in some ferromagnetic materials and many efforts have been made to understand  
the coexisting mechanism \cite{MagSup2}. In this phenomenon itinerant
electrons may be responsible, while its mechanism is not fully elucidated
yet. Many people believe that the electron Cooper pair should be $P-$
wave, since this type can be compatible with spin polarization. 
We can easily consider the similar situation in quark matter.
Since the volumes of the 
Fermi seas of quarks with different spins result in being different due
to the net presence of magnetization, we could not construct a quark
Cooper pair in a usual manner as $J^P=0^+$. Instead, we consider the 
$J^P=0^-$ pairing with orbital angular momentum $L=1$ and total
spin $S=1$. For the $S=1$ state we further consider two possibilities: 
spin-parallel pair or spin-anti-parallel pair. We first discuss the
former case in detail, which may have a direct resemblance to the electron case.
Subsequently we briefly sketch our idea about the former case. Anyway we
shall see the gap functions become anisotropic in the momentum space
like in $^3$He or nuclear matter \cite{leg,NM3P}.

We discuss another magnetic aspect in quark matter at
moderate densities, where the QCD interaction is still strong and some
non-perturbative effects still remain. One of the most important
phenomena observed there is restoration of chiral symmetry.
In the vacuum chiral symmetry is spontaneously
broken to give finite mass for quarks or nucleon; we may bear in mind such a
picture that the vacuum is in a kind of
superconducting phase with massless quark ($q$)-anti-quark ($\bar q$) pair condensate, and
the gap opened at the top of the Dirac sea corresponds to the
finite mass. As a consequence the vacuum does not possess chiral symmetry
any more. At finite densities the
suppression of $\bar qq$ excitations due to the existence of the Fermi sea
gives rise to restoration of chiral symmetry at a certain density, and
many people believe that deconfinement transition occurs at almost 
the same density.

There have been proposed 
various types of the {\it p-h} condensations at moderate densities
\cite{der0, der}, 
in which the {\it p-h} pair in scalar or tensor channel  
has the finite total momentum indicating standing waves 
(the chiral density waves). The instability for the density wave in quark matter was first discussed 
by Deryagin {\it et al.} \cite{der0} at asymptotically high densities 
where the interaction is very weak, 
and they concluded that the density-wave instability prevails over the BCS one 
in the large $N_c$ (the number of colors) limit 
due to the dynamical suppression of colored BCS pairings. 

In general, density waves are favored in 1-D (one spatial dimension) systems 
and have the wave number $Q=2k_F$ according to the Peierls instability 
\cite{kagoshima,peiel1}, 
e.g., charge density waves (CDW) in quasi-1-D metals \cite{gru1}. 
The essence of its mechanism is the nesting of Fermi surfaces 
and the level repulsion (crossing) of single particle spectra 
due to the interaction for the finite wave number. 
Thus the low dimensionality has a essential role 
to produce the density-wave states. 
In the higher dimensional systems, however, the transitions occur 
provided the interaction of a corresponding ({\it p-h}) channel is strong enough. 
For the 3-D electron gas, 
it was shown by Overhauser \cite{ove,gru} that paramagnetic state is unstable 
with respect to the formation of the static spin density wave (SDW), 
in which spectra of up- and down-spin states are deformed  
to bring about the level crossing 
due to the Fock exchange interactions, 
while the wave number does not precisely coincide with $2k_F$ 
because of the incomplete nesting in higher dimension. 

We shall see a kind of spin density wave develops there, in analogy with SDW mentioned above.. 
It occurs along with the chiral condensation 
and is represented by 
a dual standing wave in scalar and pseudo-scalar condensates 
(we have called it `dual chiral-density wave', DCDW). 
DCDW has different features in comparison with 
the previously discussed chiral density waves \cite{der0,der}. One
outstanding feature concerns its magnetic aspect; DCDW induces {\it spin
density wave}.  

\section{Ferromagnetism in QCD}

\subsection{A heuristic argument}

Quark matter bears some resemblance to electron gas interacting with the
Coulomb potential; the one 
gluon exchange (OGE) interaction in QCD has some resemblance to the Coulomb 
interaction in QED ,
and color neutrality of quark matter corresponds to total charge neutrality of
electron gas under the background of positively charged ions.

It was
Bloch who first suggested a mechanism leading to ferromagnetism of
itinerant electrons  within the Hartree-Fock approximation \cite{blo}. 
The mechanism looks very simple but largely reflects
the Fermion nature of electrons in a model-independent way. 
Since there works no direct interaction
between charged particles as a whole, the Fock exchange interaction gives a
leading contribution.
Then it is immediately conceivable that a most
attractive channel is the parallel spin pair, whereas 
the anti-parallel pair 
gives null contribution (see Eq.~(\ref{nocor}) below).
This is nothing but a
consequence of the Pauli exclusion principle: electrons with the same
spin cannot closely approach to each other, which
efficiently avoid the Coulomb repulsion. Thus a completely polarized
state is favored by the interaction. On the other hand a polarized
state should have a larger kinetic energy by rearranging the two Fermi
spheres. Thus there is a trade-off between the kinetic and interaction
energies, which leads to a {\it spontaneous spin polarization (SSP)} 
 or FM at a certain density. Subsequently it has been proved that 
Bloch's idea is qualitatively justified, but 
the critical density can not be reliably estimated without examining the
higher-order correlation diagrams \cite{yoshi,bru}; especially the ring diagrams 
have been known to
be important in the calculation of the susceptibility 
of electron gas. Recently the possibility of ferromagnetism in the
electron gas has been studied by the quantum Monte Carlo simulation and 
it has been shown that the electron gas is in ferromagnetic phase at
very low electron density \cite{cep}. Authors in ref.\cite{you} have
confirmed it experimentally.

One of the essential points we learned here is that 
we need no spin-dependent interaction in the original Lagrangian to see
SSP: a symmetry principle gives rise to a spin dependent interaction. 

Then it might be natural to ask how about in QCD. We list here some
features of QCD related to this subject. (1) the quark-gluon interaction
in QCD is rather simple, compared with the nuclear force; it is a gauge
interaction like in QED. (2) quark matter should be a color neutral
system and only the $\it Fock~ exchange$ interaction is also relevant like in
the electron system. (3) there is an additional flavor degree of freedom in
quark matter; gluon exchange never change flavor but it becomes effective through
the generalized Pauli principle. (4) quarks should be treated
relativistically, different from the electron system.

The last feature requires a new definition and formulation of SSP or FM in
relativistic systems since ``spin'' is no more a good quantum number for 
relativistic particles;
spin couples with momentum and its direction changes during the motion.  
It is well known that the Pauli-Lubanski vector $W^\mu$ is the four vector to 
represent the
spin degree of freedom in a covariant form; the spinor of the free Dirac 
equation is the eigenstate of the operator,
\beq
W\cdot a=-\frac{1}{2}\gamma_5\ds{a}\ds{k},
\label{aab}
\eeq
where a 4-axial-vector $a^\mu$ is orthogonal to $k$ s.t.
\beq
{\bf a}=\bzeta+\frac{\bk(\bzeta\cdot\bk)}{m(E_k+m)}, 
~a^0=\frac{\bk\cdot\bzeta}{m}~~~
\label{ac}
\eeq
with the axial vector $\bzeta$. We can see that $a^\mu$ is reduced to a
three 
vector $(0, \bzeta)$
in the rest frame, where we can allocate $\bzeta=(0,0,\pm 1)$ to spin ``up'' and ``down'' states.
Thus we can still use $\bzeta$ to specify the two intrinsic polarized states
even in the general Lorentz frame. To characterize the degeneracy of the plane wave solution 
$u^{(\alpha)}(k)$ ($\alpha=1,2$) for 
a positive energy state, we can use such spinors
$u^{(\alpha)}(k)$ that are eigenstates of the operator $-W\cdot a/m_q$:  
for the standard
representation of $u^{(\alpha)}(k)$ \cite{itz}, 
\beq
-\frac{W\cdot a}{m_q}u^{(\alpha)}(k)=\pm\frac{1}{2}u^{(\alpha)}(k).
\eeq
Accordingly, the polarization density matrix $\rho(k, \bzeta)$ is given by the expression,
\beq
\rho(k, \bzeta)
=\frac{1}{2m_q}(\ds{k}+m_q)P(a), \quad P(a)=\frac{1}{2}(1+\gamma_5\ds{a}),
\label{ad}
\eeq
which is normalized by the condition, ${\rm tr}\rho(k, \bzeta)=1$ \cite{ber}.

Consider the spin-polarized quark liquid with the total number density 
of quarks $n_q$
\footnote{We, hereafter, consider one flavor quark matter, since the OGE
interaction never changes flavors.}
; we denote the number densities of 
quarks with spin up and down by $n_+$ and $n_-$, respectively
, and introduce the
polarization parameter $p$ by the
equations,
$
n_\pm=\frac{1}{2}n_q(1\pm p),
$
under the condition $0\leq p\leq 1$. We assume as usual 
that three color states 
are occupied to be neutral for each momentum and spin state.
The Fermi momenta in the
spin-polarized quark matter are then 
$
k_F^\pm=k_F(1\pm p)^{1/3} 
$
with $k_F=(\pi^2n_q)^{1/3}$.
The kinetic energy density is given by the standard formula,
\beq
\E_{kin}=\frac{3}{16\pi^2}\sum_{i=\pm}\left[k_F^iE_F^i(2k_F^{i2}+m^2_q)
-m^4_q\ln\left(\frac{E_F^i+k_F^i}{m_q}\right)\right],
\label{fa}
\eeq
with the Fermi energy $E_F^i=(m^2_q+k_F^{i2})^{1/2}$.

Let us 
consider the OGE interaction between two quarks with momenta, $k$ and $q$, and 
spin vectors, $\bzeta$ and $\bzeta'$, respectively.
The color symmetric matrix element ${\cal M}^s_{{\bf k}\zeta,{\bf
q}\zeta'}$ is given only 
by the exchange term; the direct term vanishes because the color
symmetric combinations ($\sim {\rm tr}\lambda_a$) does not couple to
gluons. Thus    
\beqa
{\cal M}^s_{{\bf k}\zeta,{\bf q}\zeta'}&=&-g^2
\frac{1}{9}{\rm tr}(\lambda_a/2\lambda_a/2)\bar u^{(\zeta')}(\bq)\gamma_\mu
u^{(\zeta)}(\bp)\bar u^{(\zeta)}(\bp)\gamma^\mu u^{(\zeta')}(\bq)
\frac{-1}{(k-q)^2}
\nonumber\\
&=&\frac{4}{9}g^2\frac{1}{4}{\rm tr}
\left[\gamma_\mu\rho(k,\zeta)\gamma^\mu\rho(q,\zeta')\right]
\frac{1}{(k-q)^2},
\label{cf}
\eeqa
by the use of Eq.~(\ref{ad}). If we choose both
$\bzeta$ and $\bzeta'$ in parallel along the $z$ axis, 
$\bzeta=\bzeta'=(0,0,\pm 1)$, we have the spin-nonflip amplitude 
${\cal M}^{s, nonflip}_{\bp\bq}$, while if we choose them in anti-parallel,
$\bzeta=-\bzeta'$, we have the spin-flip amplitude 
${\cal M}^{s, flip}_{\bp\bq}$. Each form of the spin-nonflip or
spin-flip amplitude is complicated, but their average gives a
simple form,
\beq
\overline{{\cal M}}^s_{\bp\bq}=\frac{2}{9}g^2\frac{2m^2_q-k\cdot q}{(k-q)^2},
\label{cj}
\eeq
which is nothing but the matrix element for the unpolarized case \cite{bay}.
In the nonrelativistic limit, $m_q\gg |\bp|, |\bq|$, the matrix
element is reduced to the form,
\beq
{\cal M}^s_{{\bf k}\zeta,{\bf
q}\zeta'}=-\frac{2}{9}g^2\frac{m_q^2(1+\bzeta\cdot \bzeta')}{|{\bf
k}-{\bf q}|^2}, 
\label{nocor}
\eeq
so that there is {\it no correlation} between quarks with different
spins. On the other hand, there is some correlation included 
in the relativistic case.

After summing up over the
color degree of freedom and performing the integrals of the
color symmetric matrix element ${\cal M}^s_{{\bf k}\zeta,{\bf q}\zeta'}$
over the Fermi
seas of spin up and down quarks, we have the exchange energy
density $\E_{ex}$ consisting of two contributions,
\beq
\E_{ex}=\E_{ex}^{nonflip}+\E_{ex}^{flip}.
\label{da}
\eeq

In the nonrelativistic case, 
the spin-flip contribution becomes tiny and 
the dominant contribution for the OGE energy density in
Eq.(\ref{da}) comes from the spin-nonflip contribution (see Eq.~(\ref{nocor})),
\beq
\E_{ex}\sim -\frac{\alpha_ck_F^4}{2\pi^3}
\left\{(1+p)^{4/3}+(1-p)^{4/3}\right\}.
\label{fd}
\eeq
The exchange energy is negative and 
takes a minimum at $p=1$.
The form of the 
energy density (\ref{fd}) is exactly the same as in electron gas.
It is the difference of
density dependence between the contributions given 
in Eqs.~(\ref{fa}) and (\ref{fd}) which causes a
ferromagnetic instability;  
this mechanism was first 
pointed out by Bloch for electron gas \cite{blo}.

In the relativistic case there are some different features from the
nonrelativistic case. First, there is a spin-flip
contribution due to the lower component of the Dirac spinor even for
the Coulomb-like interaction. Secondly, the transverse (magnetic) gluons becomes
important, where the spin-flip effect is prominent. Finally, 
the density dependence of kinetic energy as
well as the exchange energy is very different \cite{tam}.
Before discussing the general case, we consider the relativistic
limit, $k_F^i\gg m_q$; 
the Fock exchange-energy density looks like 
\beq
\E_{ex}\sim \frac{\alpha_c}{8\pi^3}k_F^4
\left\{(1+p)^{4/3}+(1-p)^{4/3}+2(1-p^2)^{2/3}\right\}, 
\label{ff} 
\eeq
which is a decreasing function and takes
a minimum again at $p=1$. This is due to the characteristic feature 
of the spin-flip and 
spin-nonflip interactions: both give a repulsive contribution in the
relativistic limit and there is no spin-flip interaction in the
polarized state ($p=1$).
Thus, ferromagnetism in the relativistic limit arises by a different
mechanism from
that in the nonrelativistic case.

In Fig.1 a typical shape of the total energy density,
$\E_{tot}=\E_{kin}+\E_{ex}$, is depicted 
as a function of the polarisation parameter $p$, e.g. for the parameter set
,$m_q=300$MeV of the $s$ quark and 
$\alpha_c=2.2$ as in the MIT bag model \cite{deg}
\footnote{The difficulties to determine the values of these parameters 
have been discussed in ref. \cite{far}, and we must allow some range
for them.}.
We can see that paramagnetic quark matter ($p=0$) becomes unastable as
density decreases, and ferromagnetic phase is favored at a
certain density between 0.1 and 0.2 fm$^{-3}$. This phase transition is
of weakly first-order and the completely polarized ($p=1$) state appears at the
critical density.
\begin{figure}[h]
\begin{minipage}{0.49\linewidth}
\begin{center}
\includegraphics[width=7cm]{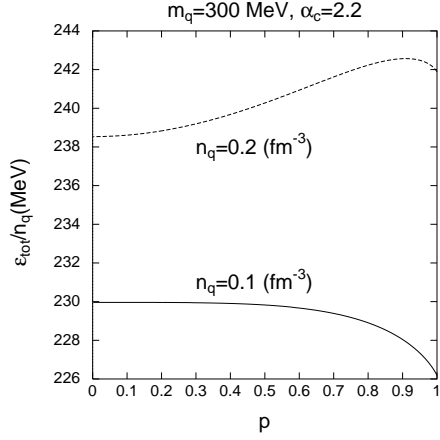}
\end{center}
\caption{First-order phase transition to the ferromagnetic state.}
\label{}
\end{minipage}
\hspace{\fill}
\begin{minipage}{0.49\linewidth}
\begin{center}
\includegraphics[width=7cm]{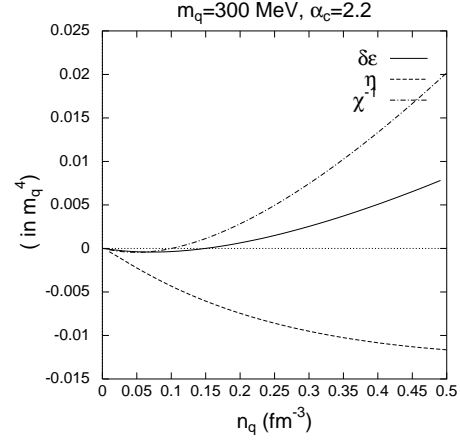}
\end{center}
\caption{Critical lines as functions of quark-number density.}
\end{minipage}
\end{figure}

To figure out the features of the ferromagnetic transition, we study
other quantities. 
For small $p\ll 1$, the energy density behaves like
\beq
\E_{tot}-\E_{tot}(p=0)=\chi^{-1} p^2+O(p^4)
\label{ha}
\eeq
with $\chi^{-1}\equiv \chi_{kin}^{-1}+\chi_{ex}^{-1}$. $\chi$ is 
proportional to the magnetic susceptibility, and its sign change
indicates a ferromagnetic transition, if it is of the second order. 
It consists of two contributions: 
the kinetic energy gives $\chi_{kin}^{-1}=k_F^5/(3\pi^2E_F)$ (c.f. the
Pauli paramagnetism), which 
changes from $\chi_{kin}^{-1}\sim O(k_F^5)$ at low densities to
$\chi_{kin}^{-1}\sim O(k_F^4)$ at high densities.
On the other hand, the Fock exchange energy gives
\beq
\chi_{ex}^{-1}=-\frac{2\alpha_ck_F^4}{9\pi^3}\left[2-\frac{k_F^2}{E_F^2}
-\frac{3m_q^2k_F}{E_F^3}\ln\left(\frac{E_F+k_F}{m_q}\right)
+\frac{4m_qk_F^2}{3E_F^2(E_F+m_q)}\right]
\label{hc}
\eeq
\cite{akh}, which is reduced to 
\beq
\chi_{ex}^{-1}\sim -\frac{4\alpha_c}{9\pi^3}k_F^4
\eeq
in the nonrelativistic limit, $p_F\ll m_q$.
In
the relativistic limit, $p_F\gg m_q$, it behaves like 
\beq
\chi_{ex}^{-1}\sim
\frac{\alpha_c}{9\pi^3}k_F^4-\frac{\alpha_c}{3\pi^3}k_F^4
=-\frac{2\alpha_c}{9\pi^3}k_F^4,
\label{hd}
\eeq
where the first term stems from the spin-nonflip
contribution, while the second term from the spin-flip contribution.
Then we can see that the effect of the spin-flip contribution
overwhelms the one of the spin-nonflip contribution. 
The interaction contribution
$\chi_{ex}^{-1}$ is always {\it negative}
, and dominant over $\chi_{kin}^{-1}$ 
at low densities,
while the kinetic contribution $\chi_{kin}$ is always {\it
positive}. If $\alpha_c>3\pi/2=4.7$, $\chi$ becomes negative over all
densities.

For a given set of $m_q$ and $\alpha_c$, $\chi$ changes its sign 
at a certain density, denoted by $n_{c1}$, and it is a signal for the
second-order phase transition.
Note that the ferromagnetic
transition in our case is of the first order, so that it is not
sufficient to only see the magnetic susceptibility; even above that density the
ferromagnetic phase may be
possible. Actually there is a range, $n_{c1}<n_q<n_{c2}$, 
where $\chi>0$ but $\E<0$.

Above the density $n_{c2}$ there is no longer the stable
ferromagnetic phase. However, the {\it metastable} state is still possible
up to the density $n_{c3}$, which is specified by the condition
s.t. $\eta\equiv\partial \E_{tot}/\partial p~|_{p=1}<0$. In Fig.2 we
depict the quantities $\chi^{-1},\delta \E$  and $\eta$ as the functions of
density, e.g. for the set
,$m_q=300$MeV and $\alpha_c=2.2$.
The crossing points with the horizontal axis indicate the critical
deisities $n_{c1}, n_{c2}$ and $n_{c3}$, respectively. 
We can see that the ferromagnetic instability occurs at low densities, 
while the metastable state can exist up to rather high densities.

Finally we show the critical lines satisfying $\chi^{-1}=0, \delta \E=0$ and 
$\eta=0$
in the QCD parameter ($\alpha_c$ and $m_q$) plane, which
seperate the three characteristic regions  
for a given density. In Fig.~3 we demonstrate them at a density $n_q=0.3$fm$^{-3}$. All the lines have the
maxima around the medium  quark mass, and the mechanism of  
ferromagnetism is different for each side of 
the maximum, as already discussed. 
If we take $m_q=300$MeV for the $s$ quark or $m_q\sim 0$MeV for the
$u$ or $d$ quark, and 
$\alpha_c=2.2$ as in the MIT bag model again \cite{deg}, the 
quark liquid can be
ferromagnetic as a metastable state.
\begin{figure}[h]
\begin{center}
\includegraphics[width=8cm]{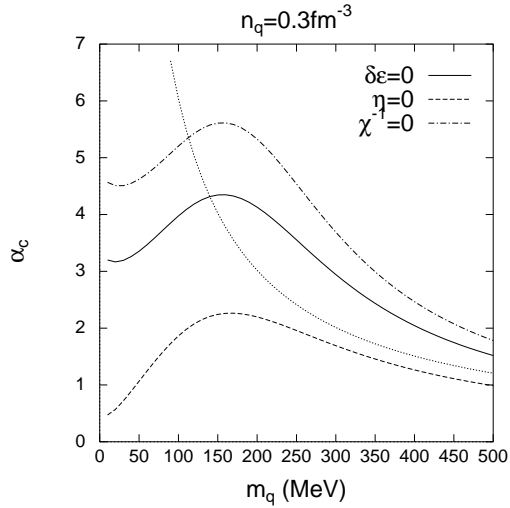}
\end{center}
\caption{Phase diagram in the coupling-strength ($\alpha_s$)-mass ($m_q$)
 plane.}
\label{}
\end{figure}

We have seen that the ferromagnetic phase is realized at low densities 
and the metastable state is plausible up to rather high densities for
a reasonable range of the QCD parameters. Our calculation is based on
the lowest-order perturbation. So we need to examine the higher-order
gluon-exchange contributions to confirm the possibility.   It should be interesting to
refer a recent paper \cite{nie}, where the author also found the
ferromagnetic transition at low densities within the perturbative
QCD calculation beyond the lowest-order diagram. 

If a ferromagnetic quark liquid exists stably or metastably around or
above nuclear density, it has
some implications on the properties of strange quark stars and strange
quark nuggets \cite{mad}. They should be magnetized in a macroscopic
scale. Considering  
a possibility to attribute magnetars to strange quark stars 
in a ferromagnetic phase,
we roughly estimate the strength of the magnetic field 
at the surface of a strange quark star. 
Taking the stellar parameters of strange quark stars 
to be similar to those for canonical neutron stars with 
the typical mass around $M_G=1.4M_\odot$, 
we find the total magnetic dipole moment $M_q$,  
$M_q=\mu_q\cdot(4\pi/3\cdot r_q^3)n_q$ for the quark sphere with
the density $n_q$ and the radius $r_q$, 
where $\mu_q$ is the magnetic moment of each quark.
 Then the dipolar magnetic field at
the star surface $r=R\simeq 10$km takes a maximal strength at the poles, 
\beq
B_{\rm
max}=\frac{8\pi}{3}\left(\frac{r_q}{R}\right)^3\mu_qn_q=10^{15}[{\rm
G}]\left(\frac{r_q}{R}\right)^3\left(\frac{\mu_q}{\mu_N}\right)
\left(\frac{n_q}{0.1{\rm fm}^{-3}}\right)
\eeq
with nuclear magneton $\mu_N$, which looks enough for magnetars.

\subsection{Self-consistent calculation}

If we understand FM or magnetic properties of quark matter more deeply,
we must proceeds to a self-consistent approach, like the Hartree-Fock
theory,  
beyond the previous perturbative argument
\footnote{Simple plane wave is the solution of the Hartree-Fock equation
in the nonrelativistic electron gas, while it is not in quark matter.}
.

We begin with an OGE action:
\beqa
    I_{int}=-g^2\frac{1}{2}\int{\rm d^4}x \int{\rm d^4}y
 \left[\bar{\psi}(x)\gamma^\mu \frac{\lambda_a}{2} \psi(x)\right]
D_{\mu \nu}(x,y)
 \left[\bar{\psi}(y)\gamma^\nu \frac{\lambda_a}{2} \psi(y)\right], 
\label{ogeaction}
\eeqa
where $D^{\mu\nu}$ denotes the gluon propagator. 
By way of the mean-field approximation, we have 
\beq
 I_{MF}=\int \frac{{\rm d}^4 p}{(2 \pi)^4} 
                \bar\psi(p) G_A^{-1}(p) \psi(p). 
\label{mfield}
\eeq
The inverse quark Green function $G_A^{-1}(p)$ involves various self-energy 
(mean-field) terms, of which we only keep the color singlet
particle-hole mean-field 
$V(p)$,
\beq
G_A(p)^{-1}= \sla{p}-m+\sla{\mu}+V(p).
\label{gainv}
\eeq
Taking into account the lowest diagram, we can then write down the 
self-consistent equations for the mean-field, $V$:
\beqa
-V(k)=(-ig)^2 \int \frac{{\rm d}^4p}{i(2\pi)^4} \{-iD^{\mu \nu}(k-p)\} 
      \underbrace{\gamma_\mu \frac{\lambda_\alpha}{2} \{-iG_A(p)\} 
      \gamma_\nu \frac{\lambda_\alpha}{2}}_{(A)}  
\label{self1}.
 \eeqa

Applying the Fierz transformation for the OGE action (\ref{ogeaction})
we can see that 
there appear the color-singlet scalar, pseudo-scalar, vector and axial-vector
self-energies by the Fock exchange interaction. Taking the Feynman gauge
for the gluon propagator, a manupilation gives   
\beqa
     (A)\!\!\!&=&\!\!\!\frac{N_c^2-1}{4N_c^2}\frac{1}{N_f}\!\!\left\{{\rm
      Tr}(G_A)\!+\!i\gamma_5{\rm
      Tr}(G_Ai\gamma_5)\!-\!\frac{1}{2}[\gamma^\mu{\rm
      Tr}(G_A\gamma_\mu)\!+\!\gamma_5\gamma^\mu{\rm Tr}(G_A\gamma_5\gamma_\mu)]\!\!\right\}\nonumber\\
&+&\{{\rm color~ non\!{\rm-}\!singlet~ or~ flavor~ non\!{\rm-}\!singlet~ terms}\}.
\eeqa
When we restrict the ground state to be an eigenstate with respect to
color and flavor, there is only left the first term which is color
singlet and flavor singlet. Still we  
must take into account various mean-fields in $V$,
$V=U_s+i\gamma_5U_{ps}+\gamma_\mu U_v^\mu+\gamma_\mu\gamma_5U_{av}^\mu$
with the mean-fields $U_i$. Here we only retain ${\bf U}_{av}(\equiv {\bf U}_{A})$ for
simplicity and suppose that others to be vanished; 
\begin{equation}
V(k) = \bgamma \gamma_5 \cdot {\bf U}_A({\bf k}), 
\label{av}
\end{equation}
with the static axial-vector mean-field $U_A({\bf k})$. 

The poles of $G_A(p)$, $\det$$G^{-1}_A$($p_0+\mu$$=$$\epsilon_n$)$=$$0$,
give the single-particle energy spectrum:
\beqa
&& \epsilon_n=\pm \epsilon_\pm \\
  && \epsilon_{\pm}({\bf p})= \sqrt{{\bf p}^2+{\bf U}_A^2({\bf p})+m^2 \pm 2 
                        \sqrt{m^2 {\bf U}_A^2({\bf p})+({\bf p}\cdot {\bf U}_A({\bf p}))^2 }},
\label{eig}
\eeqa
where the subscript in $\epsilon_s({\bf p}), s=\pm$ represents spin degrees of
freedom, and the dissolution of the degeneracy
corresponds to the {\it exchange splitting} of 
different ``spin'' states; the spectrum is reduced to a familiar form
$ \epsilon_{\pm}\sim m+\frac{p^2}{2m}\pm |{\bf U}_A|$ in the
non-relativistic limit \cite{yoshi}.

\begin{figure}[h]
\begin{center}
\includegraphics[width=10cm]{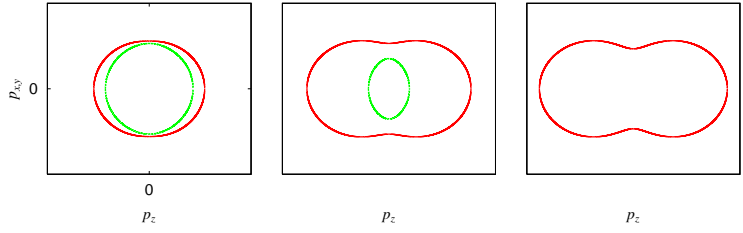}
\end{center}
\caption{Modification of the Fermi sea as $U_A$(=const.) 
is increased from left
 to right. The larger Fermi sea ($F^-$) takes a prolate shape, while
 the smaller one ($F^+$) an oblate shape for a given $U_A$. In the
 large $U_A$ limit (completely polarized case), $F^+$ disappears as 
in the right panel.}
\label{multi}
\end{figure}

There appear two Fermi seas with different volumes for a given quark number 
due to the exchange splitting in the energy spectrum. 
The appearance of the rotation symmetry breaking term, $\propto {\bf
p}\cdot {\bf U}_A$ in the energy
spectrum implies deformation of the Fermi sea: thus 
rotation symmetry is violated in the momentum space as well as the
coordinate space, $O(3)\rightarrow O(2)$. Accordingly the Fermi sea
of majority quarks exhibits a ``prolate'' shape ($ F^-$), while that 
of minority quarks an ``oblate'' shape ($F^+$) as seen in Fig.~\ref{multi}.

Then the self-consistent equation (\ref{self1}) is reduced to the form,
\beq
  U_A({\bf k})=\frac{N_c^2-1}{4N_c}{g}^2 \int \frac{{\rm d}^3
  p}{(2\pi)^3}\sum_{s=\pm}\frac{1}{\epsilon_s(\vp)^2-|{\bf k}-{\bf p}|^2} 
\theta(\mu-\epsilon_s(\vp))
\frac{U_A({\bf p}) +s \beta_p}{\epsilon_s(\vp)}
\label{UA1s}
\eeq 
with $\beta_p=\sqrt{p_z^2+m^2}$ by taking ${\bf U}_A$ along the
$z$ axis. Here we have discarded the contribution of the Dirac seas and
only taken into account that of the Fermi seas.

In the following we demonstrate some
numerical results by replacing the original OGE
by the ``contact'' (zero-range) interaction,
$
D^{\mu\nu}\rightarrow -g^{\mu\nu}/\Lambda^2,
$
which may correspond to the Stoner model in the condensed matter physics
\cite{yoshi}.
\footnote{When we take into account the Debye screening, the time
component of the gluon propagator becomes finite range due to the Debye
screening. If typical momentum transfer $Q$ is much smaller than 
the screening mass $M_D^2 \sim N_f g^2
\mu^2/(2\pi^2)$ \cite{LeBe}, we may replace the OGE with the infinite
range by the zero-range effective interaction.}   

We can easily see that the mean-field $U_A$ becomes then
momentum-independent, 
and the expression for $U_A$,
   Eq.~(\ref{UA1s}), is proportional to the simple sum of the expectation
   value of the spin operator over the Fermi seas;
\beqa
{\bar s_z}=\frac{1}{2}\langle
\Sigma_z\rangle&=&-i\int_C\frac{d^4p}{(2\pi)^4}{\rm tr}
\gamma_5\gamma_3G_A(p)\nonumber\\
&=&\frac{1}{2}\left[\int_{F^+}\frac{d^3p}{(2\pi)^3} 
\frac{U_A({\bf p})+\beta_p}{\epsilon_+({\bf p})}+\int_{F^-}\frac{d^3p}{(2\pi)^3} 
\frac{U_A({\bf p})-\beta_p}{\epsilon_-({\bf p})}\right].
\label{magnet}
\eeqa

\subsection{Phase diagram on the temperature-density plane} 
We will present the phase diagram in the three-flavor case under 
two conditions \cite{nak05}:
the chemical equilibrium condition (CEC) $\mu_u=\mu_d=\mu_s$ and 
the charge neutral condition without electrons (CNC) $\rho_u=\rho_d=\rho_s$, 
where quark masses are taken as $m_u=m_d=5$MeV and $m_s=150-350$MeV, 
i.e., $\mu_{s}=\sqrt{\mu_{u,d}^2+m_s^2-m_{u,d}^2}$ for $T=0$.  
In both conditions, 
since the spin polarization caused by the axial-vector mean-field is
fully enhanced by the quark mass for given density or temperature,   
choice of the current quark mass seriously affects the results;  
especially, largeness of the strange quark mass has an essential effect on spin polarization. 
To get the phase diagram or critical line on the temperature-density plane, 
we use the thermodynamic potential $\Omega$ within the mean-field approximation, 
\begin{eqnarray}
\Omega&=&
-N_c \sum_{B=\pm1}~\sum_{s=\pm}~ \sum_{i=u,d,s} \idk 
 T\log
\left\{ \exp\left[ -\frac{\epsilon_s(\bk,m_i,U_A)-B\mu_i}{T}\right]+1 \right\}
\nn
&&-N_c \sum_{s=\pm}~ \sum_{i=u,d,s} \idk \epsilon_s(\bk,m_i,U_A) +\frac{U_A^2}{4 \tilde{g}^2},
\label{PD}
\end{eqnarray}
where we have used the ``contact'' interaction, $\tilde g^2\equiv g^2/\Lambda^2$, in
place of the OGE interaction.
Note that we take into accout the vacuum contribution in this formula
(the second term in Eq.~(\ref{PD})), which should  be regularized by
,e.g., the proper-time method (see \S 4.3). 
We can confirm that the thermodynamic potential reproduces the self-consistent
equation for the order parameter $U_A$  
Eq.~(\ref{UA1s}) in the three-flavor case, except the vacuum
contribution. The vacuum (the Dirac sea) contribution always works against spin
polarization as it should do, while the contribution of the Fermi sea
gives rise to spontaneous spin polarization.

\begin{figure}[h]
\begin{center}
\includegraphics[width=8cm]{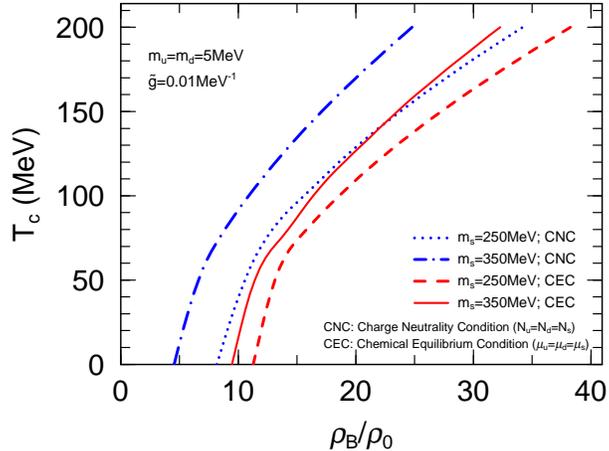}
\end{center}
\caption{Phase diagram for two cases of strange quark masses, $250, 350$MeV.}
\label{pd3}
\end{figure}
Fig.~\ref{pd3} shows the critical temperature (the Curie temperature) as a function of
baryon-number density under the two conditions mentioned above; 
CEC and CNC. 
We can see that CNC tends to facilitate the system having spin polarization than CEC.  
This is because CNC holds the larger strange-quark density than CEC. 

\begin{figure}
\begin{center}
\includegraphics[width=8cm]{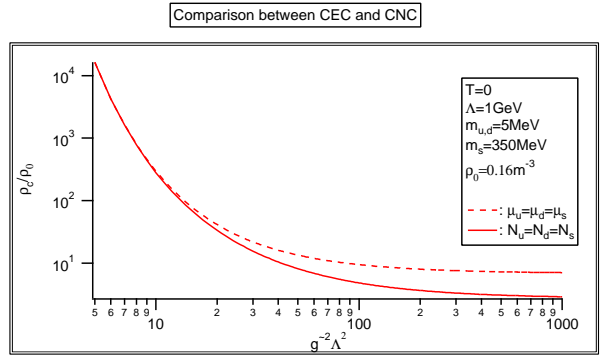}
\end{center}
\caption{Critical density as a function of $\tilde{g}$ under two conditions: CEC and CNC.}
\label{comp1}
\end{figure}

Since the axial-vector mean-field arises from the Fock exchange
interaction among quarks in the Fermi sea  
and causes a kind of particle-hole condensation,   
there exists a critical density for a given coupling constant $\tilde{g}$.   
We show the critical density by varying the effective coupling constant $\tilde{g}$ in Fig.~\ref{comp1}.  
The critical density is more lowered with the larger coupling strength, 
and this tendency is remarkable in the case of CNC. 
The result also indicates that even for the weak-coupling regime in QCD,  
spin polarization may appear at sufficiently large densities and low temperatures.    

\section{Color magnetic superconductivity}
\subsection{General framework}

If FM is realized in quark matter, it might be in the CSC phase.
In this section we discuss a possibility of the coexistence of FM and
CSC, which we call {\it Color magnetic superconductivity} \cite{nak}. 

Recall the OGE action Eq.~(\ref{ogeaction}). 
By way of the mean-field approximation, we have 
\beq
 I_{MF}=\frac{1}{2} \int \frac{{\rm d}^4 p}{(2 \pi)^4} 
                \left( \begin{array}{l} 
                          \bar{\psi}(p)   \\
                          \bar{\psi}_c(p) \\
                       \end{array} \right)^T
                  G^{-1}(p)
                \left( \begin{array}{l} 
                          \psi(p)   \\
                          \psi_c(p) \\
                       \end{array} \right) \\
\label{mfield}
\eeq
in the Nambu-Gorkov formalism, allowing not only the
particle-hole but also the particle-particle mean-field.
The inverse quark Green function $G^{-1}(p)$ involves various self-energy 
(mean-field) terms, of which we only keep the color singlet particle-hole 
$V(p)$ and color $\bar 3$ particle-particle ($\Delta$) mean-fields; 
the former is responsible to 
ferromagnetism, while the latter to superconductivity,
\beqa
G^{-1}(p)&=&\left( \begin{array}{cc}
                      \sla{p}-m+\sla{\mu}+V(p) & 
                      \gamma_0 \Delta^\dagger(p) \gamma_0  \\
                      \Delta(p) & 
                      \sla{p}-m-\sla{\mu}+\overline{V}(p) \\
                          \end{array} \right),\nonumber\\
         &=&\left( \begin{array}{cc}
                            G_{11}(p) & G_{12}(p) \\
                            G_{21}(p) & G_{22}(p) \\
                            \end{array} \right)^{-1} \label{fullg2} 
\eeqa
where
\beq
\psi_c(k) = C \bar{\psi}^T(-k),~~~\overline{V} \equiv C V^T C^{-1}. 
\eeq
Taking into account the lowest diagram, we can then write down the 
self-consistent equations for the mean-fields, $V$ and $\Delta$:
\beqa
-V(k)=(-ig)^2 \int \frac{{\rm d}^4p}{i(2\pi)^4} \{-iD^{\mu \nu}(k-p)\} 
      \gamma_\mu \frac{\lambda_\alpha}{2} \{-iG_{11}(p)\} 
      \gamma_\nu \frac{\lambda_\alpha}{2}  
\label{self1s}.
 \eeqa
and
\beq
  -\Delta(k)=(-ig)^2 \int \frac{{\rm d}^4p}{i(2 \pi)^4} \{-iD^{\mu \nu}(k-p)\}
               \gamma_\mu \frac{-(\lambda_\alpha)^T}{2}
                 \{-iG_{21}(p) \} 
               \gamma_\nu \frac{\lambda_\alpha}{2},  
\label{gap1} 
\eeq
(c.f. Eq.~(\ref{self1})).
\begin{figure}[h]
\begin{center}
\includegraphics[width=5cm]{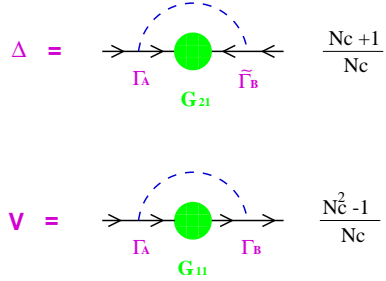}
\end{center}
\caption{Graphical interpretations of the coupled equations (\ref{self1s}) and
 (\ref{gap1}) with coefficients in front of R.H.S. given by $N_c$. The
 lower diagram becomes dominant in the large $N_c$ limit.}
\end{figure}

The structure of Eq.~(\ref{self1s}) is the same as Eq.~(\ref{self1}),
and we can see 
 there appear the color-singlet scalar, pseudoscalar, vector and axial-vector
self-energies by applying the Fierz transformation .   
Here we retain only $U_s, U_v^0, U_{av}^3$
in $V$ and suppose that others to be vanished as before.
We shall see this ansatz gives 
self-consistent solutions for Eq.(\ref{self1s}) within the zero-range
approximation for the OGE interaction because of axial and reflection
symmetries of the Fermi seas. We furthermore discard the scalar mean-field $U_s$ and 
the time component of the vector mean-field $U_v^0$ for simplicity 
since they are irrelevant for the spin degree of freedom.

According to the above assumptions and considerations 
the mean-field $V$ in Eq.(\ref{fullg2}) renders 
\begin{equation}
V = \gamma_3 \gamma_5 U_A, ~~~U_A\equiv U_{av}^3 , 
\end{equation}
with the axial-vector mean-field $U_A$, as in Eq.~(\ref{av}).
Then the diagonal component of the Green function $G_{11}(p)$ is written as
\begin{equation}
  G_{11}(p)=\left[ G_A^{-1}-
              \gamma_0 \Delta^\dagger \gamma_0 \tilde{G}_A \Delta \right]^{-1} 
\end{equation}
with
\begin{eqnarray}
   G_A^{-1}(p) &=& \sla{p}-m+\sla{\mu}-\gamma_5 \gamma_3 U_A, \\
  \tilde{G}_A^{-1}(p) &=& \sla{p}-m-\sla{\mu}-\overline{\gamma_5 \gamma_3} U_A,
\end{eqnarray}
where  $\overline{\gamma_5 \gamma_3}=\gamma_5 \gamma_3$ and 
$G_A(p)$ is the Green function with $U_A$  
which is determined self-consistently by way of Eq.~(\ref{self1}).  

\subsection{$^3P$ type anisotropic pairing}

Before constructing the gap function $\Delta$, 
we first find the single-particle spectrum 
and their eigenspinors in the absence of $\Delta$, which 
is achieved by diagonalization of the operator $G_A^{-1}$. We have
already known four single-particle energies 
$\epsilon_\pm$ (positive energies) and $-\epsilon_\pm$ (negative energies), 
which are given as 
\begin{eqnarray}
&& \epsilon_{\pm}({\vp}) = 
  \sqrt{{\vp}^2 + U_A^2 + m^2 \pm
             2 U_A \sqrt{m^2 + p_z^2 }}, 
\label{eig}
\end{eqnarray}
and the eigenspinors $\phi_s,~s=\pm $ should satisfy the equation, 
$G_A^{-1}(\epsilon_s, {\bf p})\phi_s=0$. 

Here we take the following ansatz for $\Delta$:  
\begin{eqnarray}
\Delta({\vp})&=&\sum_{s=\pm} \tilde{\Delta}_s({\vp}) B_s({\vp}),\nonumber\\
B_s({\vp})&=&\gamma_0 \phi_{-s}({\vp}) \phi_{s}^\dagger({\vp}).
\label{delta}
\end{eqnarray} 

\begin{figure}[h]
\begin{minipage}{0.49\linewidth}
\begin{center}
\includegraphics[width=5cm]{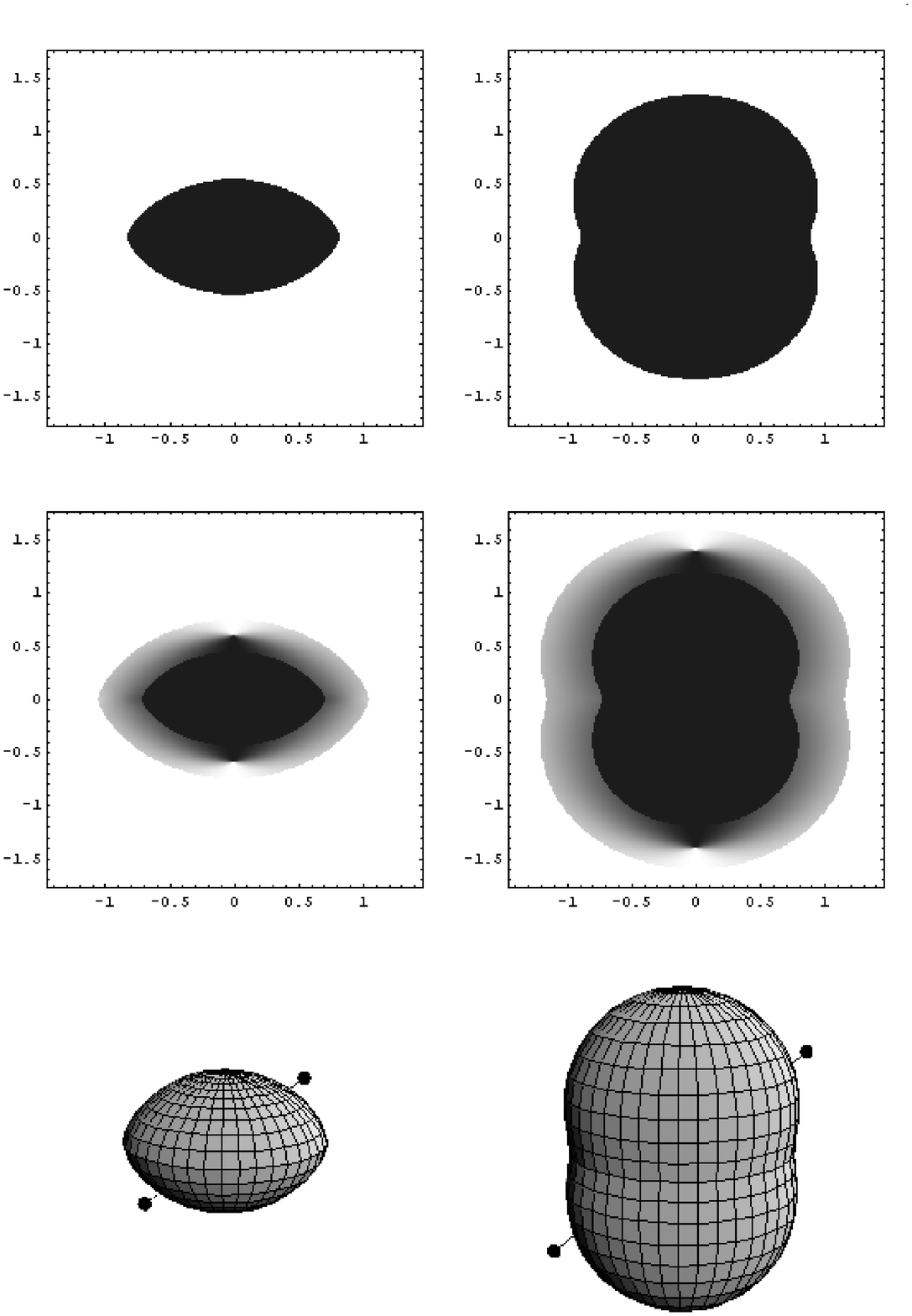}
\end{center}
\caption{Deformed Fermi seas and the quark pair on each surface. The top
 figures show those in the absence of $\Delta_\pm$ and the middle
 figures  diffusion of the Fermi surfaces in the presence of 
 $\Delta_\pm$. The bottom ones show the quark pairing on the Fermi
 surfaces.}
\label{deff}
\end{minipage}
\hspace{\fill}
\begin{minipage}{0.49\linewidth}
\begin{center}
\includegraphics[width=5cm]{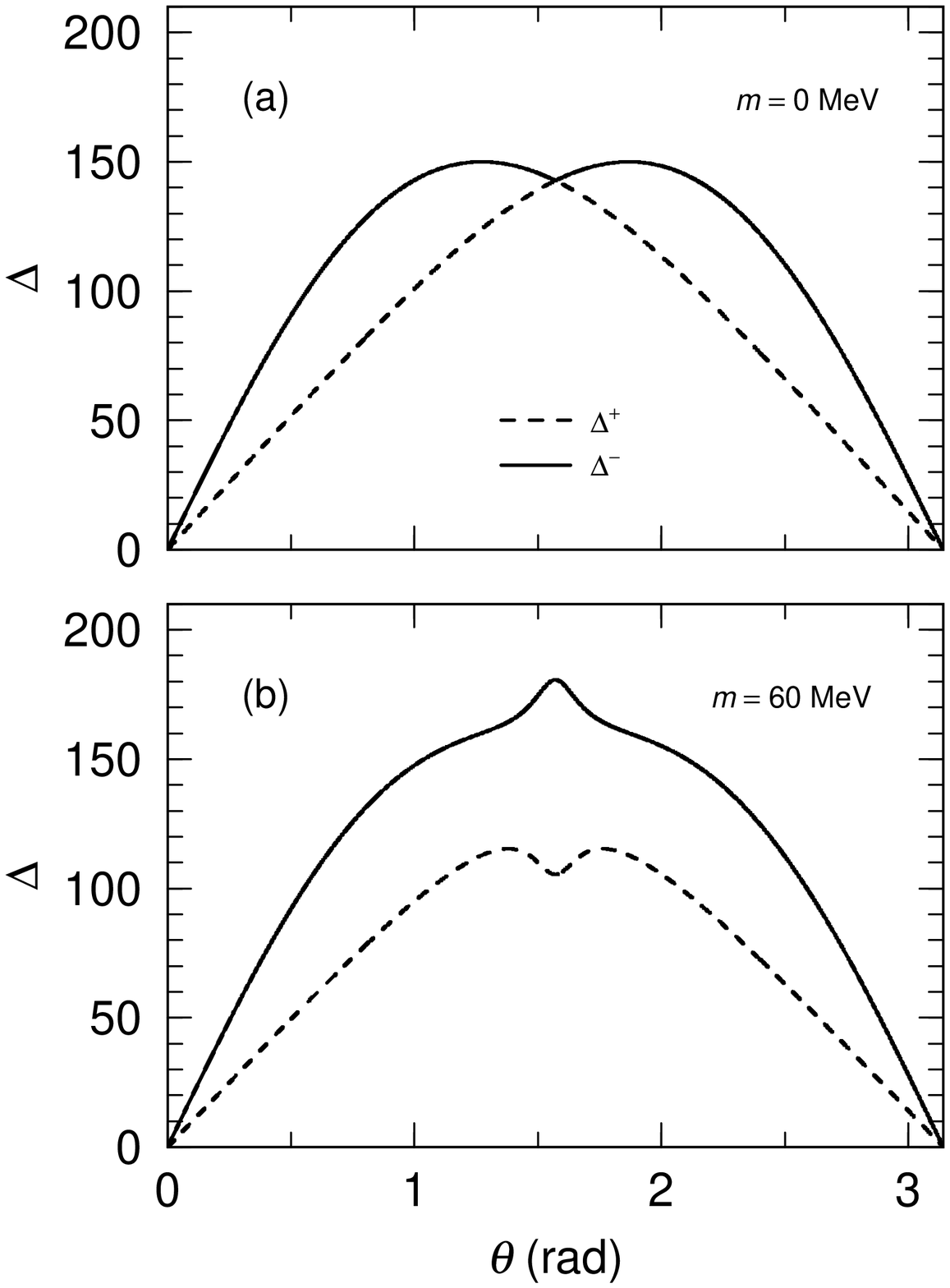}
\end{center}
\caption{Schematic view of the polar-angle dependence of the gap
 functions at the Fermi surface, (a) for $m=0$ and (b) for $m\neq 0$.}
\label{gapf}
\end{minipage}
\end{figure}

The structure of the gap
function (\ref{delta}) is then inspired by a physical consideration of a
 quark pair as in the usual BCS theory: 
we consider here the quark pair on each Fermi surface 
with opposite momenta, ${\bf p}$ and $-{\bf p}$ so that they result in a linear combination of
$J^\pi=0^-, 1^-$ (see Fig.~\ref{deff})
\footnote{Recently spin-one color superconductivity has been also studied in
the normal matter \cite{scha}.}
.

 $\tilde\Delta_s$ is still a matrix in the color-flavor space. Since  
the anti-symmetric nature of the fermion self-energy imposes a constraint 
on the gap function \cite{bai},
\begin{eqnarray}
C \Delta({\vp}) C^{-1}=\Delta^T({-\vp}).
\end{eqnarray}
$\tilde{\Delta}_n({\vp})$ must be a symmetric matrix 
in the spaces of internal degrees of freedom. 
Taking into account the property that the most attractive channel of 
the OGE interaction is  
the color anti-symmetric ${\bar 3}$ state, it must be in the flavor singlet 
state.
Thus we can  choose the form of the gap function as 
\beq
\left(\tilde\Delta_s\right)_{\alpha\beta;ij}=\epsilon^{\alpha\beta 3}\epsilon^{ij}\Delta_s
\eeq
for the two-flavor case (2SC), where $\alpha,\beta$ denote the color indices and
$i,j$ the flavor indices. 
Then the quasi-particle spectrum can be obtained by
looking for poles of the diagonal Green function, $G_{11}$: 
\beqa
E_{s}({\vp})&&=\left\{
 \begin{array}{ll}
 \sqrt{(\epsilon_s({\vp})-\mu)^2+|\Delta_s({\vp})|^2} & \mbox{for color 1, 2} \\
 \sqrt{(\epsilon_s({\vp})-\mu)^2}            & \mbox{for color 3} 
 \end{array} 
        \right.  
\label{qusiE}
\eeqa
Note that the quasi-particle energy is independent of color and flavor
in this case, since we have assumed a singlet pair in flavor and color. 

Gathering all these stuffs to put them in the self-consistent
equations, we have the coupled gap equations for $\Delta_s$,
\beqa
 \Delta_{s'}(k,\theta_k)\!=\!\frac{N_c\!+\!1}{2N_c} \tilde{g}^2 
 \!\!\int\!\! \frac{{\rm d}p\, {\rm d}\theta_p}{(2\pi)^2} p^2 \sin\theta_p
   \!\!\sum_s T_{s' s}(k,\theta_k,p,\theta_p) 
   \frac{\Delta_s(p,\theta_p)}{2 E_s(p,\theta_p)}, \label{GAP1}  
\eeqa
and the equation for $U_A$,
\beq
  U_A=-\frac{N_c^2-1}{4N_c^2}\tilde{g}^2 \int \frac{{\rm d}^3 p}{(2\pi)^3} 
\sum_s\left[\theta(\mu-\epsilon_s(\vp))+2v_s^2({\vp})\right]
\frac{U_A +s \beta_p}{\epsilon_s(\vp)}, 
\label{UA1} 
\eeq
within the ``contact'' interaction, $\tilde g^2\equiv g^2/\Lambda^2$,
   where $v_s^2(\vp)$ denotes the momentum distribution of the
   quasi-particles. We find that the expression for $U_A$,
   Eq.~(\ref{UA1}), is nothing but the simple sum of the expectation
   value of the spin operator with the weight of the occupation
   probability of the quasi-particles $v_s^2$ for two colors and the
   step function for remaining one color (cf. (\ref{UA1s})). 

Carefully analyzing the structure of the function $T_{s's}$ in
   Eq.~(\ref{GAP1}),
 we can easily find that the gap function
   $\Delta_s$ should have the polar angle ($\theta$) dependence on the Fermi
   surface,
\beq
\Delta_s(p^F_s,\theta)=\frac{p^F_s(\theta)\sin \theta}{\mu}
\left[ -s\frac{m}{\sqrt{m^2+( p^F_s(\theta) \cos \theta )^2}} R + F \right],
\label{polar}
\eeq
with constants $F$ and $R$ to be determined (see Fig.~\ref{gapf}).

As a characteristic feature, both the gap functions have nodes at poles
($\theta=0,\pi$) and take the maximal values at the vicinity of equator
($\theta=\pi/2$), keeping the relation, $\Delta_- \geq \Delta_+
$. This feature is very similar to $^3 P$ pairing in liquid $^3$He or
nuclear matter \cite{leg,NM3P}; actually we can see our pairing function
 Eq.~(\ref{polar}) to exhibit an 
effective $P$ wave nature by a genuine relativistic effect by the Dirac
spinors \cite{alf3}.
Accordingly the quasi-particle distribution is diffused (see Fig.~\ref{deff})
 
We demonstrate some self-consistent solutions here.
Since we have little information to determine the values of the parameters 
$\tilde{g}$ and $\delta$ 
(there may be other more reasonable form factors than the present cut-off
function), and 
our purpose is to figure out qualitative properties of spin polarization
in the color superconducting phase,  we mainly set in the following calculations 
them as $\tilde{g}=0.13$ MeV$^{-1}$ and $\delta=0.1\mu$, for example, 
which is not so far from the couplings in NJL-like models \cite{CSC3,Ripka,nam}.

\begin{figure}[h]
\begin{center}
\includegraphics[width=5cm]{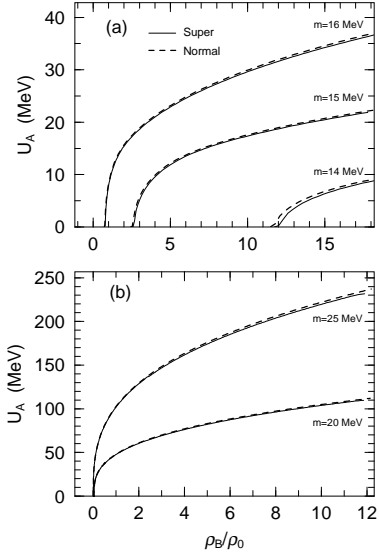}
\end{center}
\caption{Axial-vector mean-field as a function of baryon number density
 $\rho_B$ ($\rho_0=0.16$ fm$^{-3}$) 
for $\tilde{g}=0.13$ MeV$^{-1}$ and $\delta=0.1\mu$. 
{\bf (a)} for $m=14 \sim 16$ MeV and {\bf (b)} for $m=20$ and $25$ MeV.
Dashed (Solid) lines are obtained in the normal (color superconducting) phase.}
\label{UA}
\end{figure}

We first examine spin polarization in the absence of CSC.
In Fig.~\ref{UA}   
we show the the axial-vector mean-field $U_A$, 
with $\Delta_\pm $ being set to be zero,
as a function of baryon number density $\rho_B (\equiv \rho_q /3)$ 
relative to the normal nuclear density $\rho_0=0.16$ fm$^{-3}$ 
for $m=14 \sim 25$ MeV (dashed lines).
It is seen that the axial-vector mean-field (spin polarization) 
appears above a critical density and
becomes larger as baryon number density gets higher.
Moreover, 
the results for different values of the quark mass show that
spin polarization grows more for the larger quark mass.
This is because a large quark mass gives rise to much difference 
in the Fermi seas of two opposite ``spin'' states, 
which leads to growth of the exchange energy in the axial-vector channel. 

Next we solve the coupled equations (\ref{GAP1}) and (\ref{UA1}) with Eq.~(\ref{polar}).
Results for $U_A$, $R$ and $F$  are shown in 
Fig.~\ref{UA} (solid lines) 

As a consequence, we can say that FM and CSC barely interfere with each
other \cite{nak}.

\subsection{Another possibility - Gapless type pairing}

Nowadays there have been many studies about the pairing of quarks in the
two Fermi spheres with different sizes, which is caused by the mass and
charge differences among three-flavor quarks. It is well known that fermion
pairing between two different Fermi surfaces gives rise to the LOFF
phase \cite{ful,cas} or the gapless superconducting phase
\cite{BCS,alf2,tin}. There have been discussions about the phase
separation and the mixed phase in these context \cite{bed,sho}.

In the presence of magnetization we have seen that there are two
Fermi seas with different size and deformation, depending on the 
spin polarization. So we can consider
another pairing than the previous one: two quarks with opposite 
momenta and polarizations with each other take part in the pairing 
\cite{naw}. Introducing the following notation,
\beq
\epsilon_n=\{\epsilon_-,\epsilon_+,-\epsilon_-\epsilon_+\}~~~(n=1\sim 4)
\eeq
for the single quark energy by using Eq.~(\ref{eig}).
The pairing function can be written in the similar form to Eq.~(\ref{delta})
\beq
\Delta(\vp)=\sum_{n=1}^4\tilde\Delta_n(\vp)B_n(\vp)
\eeq 
with 
\beq
B_n(\vp)=\gamma_0\phi_{-\tilde n}(\vp)\phi^\dagger_n(\vp),
\eeq
where $\phi_{-\tilde n}(\vp)$ is defined by $\phi_{-\tilde n}(\vp)\equiv
\phi_{-1+(-1)^n}(\vp)$. As a combination of the quark pair in the color
and flavor spaces, 
we assume it to be anti-symmetric in both spaces,
\beq
\left[\Delta_n(\vp)\right]_{\alpha\beta;ij}=\epsilon_{\alpha\beta 3}\epsilon_{ij}\Delta_n(\vp)
\eeq
as before. Then the quasi-particle energy is given by 
\beqa
E_{n}({\vp})_\pm&&=\left\{
 \begin{array}{ll}
 E_n^A\pm\sqrt{(E_n^S)^2+|\Delta_n({\vp})|^2} & \mbox{for color 1, 2} \\
 \pm\sqrt{(\epsilon_n({\vp})-\mu)^2}            & \mbox{for color 3} 
 \end{array} 
        \right.  
\label{qusiE2}
\eeqa
with 
\beq
E_n^{S,A}=\frac{(\epsilon_n-\mu)\mp(\epsilon_{\tilde n}-\mu)}{2},
\eeq
which clearly exhibits a gapless excitation. We can also see that the
gap function shows $\cos\theta$-like dependence on the Cooper surface 
defined by the equation, $E_n^S(p,\theta)=0$. These features resemble
those given in ref.\cite{kar}, where the electron pairing with spin
anti-parallel component of the $S=1$ triplet is
considered in the presence of magnetization.

\section{Dual chiral density wave}
\subsection{Chiral symmetry restoration and 
Instability of the directional mode}

We consider here another type of magnetism in quark matter at moderate
densities, which is closely connected with chiral symmetry. We shall see
that the ground state in the spontaneously symmetry breaking (SSB) phase 
becomes unstable with respect to
producing a density wave. Accordingly the quark magnetic moment
spatially oscillates and a kind of spin density wave is induced.
The density wave can be described as a dual
standing wave in the scalar and pseudo-scalar densities \cite{tat04},
where they spatially oscillate in the phase difference of $\pi/2$ to
each other.
It is well known that chiral symmetry is spontaneously broken  
due to the quark ($q$)-anti-quark ($\bar q$) pair condensate 
in the vacuum and at low densities; since we take the vacuum as an
eigenstate of parity operation, only the scalar density is non vanishing 
to generate finite mass of quarks.
Geometrically both the scalar and pseudo-scalar densities always 
reside on the chiral sphere with the finite modulus in the SSB phase, and any chiral 
transformation with a constant chiral angle $\theta_a$ shifts each value on the sphere, 
leaving the QCD Lagrangian invariant. 
The spatially variant chiral angle $\theta({\bf r})$ 
represents the degree of freedom 
of the Nambu-Goldstone mode in the SSB 
vacuum. The dual chiral density wave (DCDW) 
is described by such a chiral angle 
$\theta({\bf r})$.
When the chiral angle has some space-time 
dependence, there should appear extra terms in the effective potential
as a consequence of chiral symmetry:
one trivial term
is the one describing the quark and DCDW   
interaction due to 
the non-commutability of $\theta({\bf r})$ with the kinetic
(differential) operator in the Dirac operator.  
Another one is nontrivial and comes from the 
vacuum polarization effect: the energy spectrum of the quark is modified in the
presence of $\theta({\bf r})$ and thereby the vacuum energy has an 
additional term, $\propto (\nabla\theta)^2$ in the lowest order. 
This can be regarded as an appearance of the kinetic term for
DCDW through the vacuum polarization \cite{sug}.  Thus, the interaction
becomes strong enough to overwhelm the kinetic energy increase, the
state becomes unstable to generate DCDW.

\begin{figure}[h]
\begin{center}
\includegraphics[width=5cm,keepaspectratio]{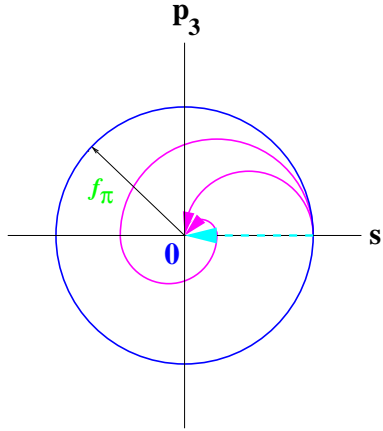}
\end{center}
 \caption{Schematic view of the possible paths of symmetry restoration
 in the chiral space. There may be possible another paths utilizing the
 degree of freedom of the chiral angle, besides the usual one along $\theta=0$.}
\label{spiral}
\end{figure}

Many studies have suggested that chiral symmetry is restored at a certain
density by suppression of $q\bar q$ excitation due to the presence
of the Fermi sea, where none of the mean-fields is present. 
In usual discussion of such symmetry restoration, one implicitly
discards the pseudo-scalar mean-field and is concentrated in the behavior
of the scalar mean-field, while there is no compelling
reason for the pseudo-scalar density to be vanished. Allowance of the
degree of freedom of the chiral angle is nothing else but 
the appearance of DCDW. Thus we can say that instability of the ground
state with respect to forming DCDW 
provides another path to symmetry restoration (see Fig.~\ref{spiral}).

\subsection{DCDW in the NJL model}

Taking the Nambu-Jona-Lasinio (NJL) model as a simple but nontrivial 
example, we explicitly demonstrate that quark matter 
becomes unstable for a formation of DCDW above 
a certain  
density; the NJL model has been  
recently used as 
an effective model of QCD, embodying spontaneous breaking of chiral symmetry 
in terms of quark degree of freedom \cite{kle}
\footnote{We can see that the OGE interaction gives the same 
form after the Fierz transformation in the zero-range limit \cite{nak}}
.
We shall explicitly see the DCDW state exhibits a ferromagnetic property.

We start with the NJL Lagrangian with $N_f=2$ flavors and $N_c=3$ colors,
\beq
{\cal L}_{NJL}
=\bar\psi(i\ds{\partial}-m_c)\psi+G[(\bar\psi\psi)^2+
(\bar\psi i\gamma_5\mbox{\boldmath$\tau$}\psi)^2],
\label{njl}
\eeq
where $m_c$ is the current mass, $m_c\simeq 5$MeV.
Under the Hartree approximation, we linearize Eq.~(\ref{njl}) by partially 
replacing the bilinear quark fields by their expectation values 
with respect to the ground state. 

In the 
usual treatment to  
study the restoration of chiral symmetry at finite density,  
authors implicitly discarded the pseudo-scalar mean-field, while  
this is justified only for the vacuum of a definite parity.
We assume here the following mean-fields,
\beqa
\langle\bar\psi\psi\rangle&=&\Delta\cos(\bf q\cdot\bf r) \nonumber\\
\langle\bar\psi i\gamma_5\tau_3\psi\rangle&=&\Delta\sin(\bf q\cdot\bf r), 
\label{chiral}
\eeqa
and others vanish
\footnote{It would be interesting to see that the DCDW  
configuration is similar to pion condensation in high-density nuclear 
matter within the $\sigma$ model, 
considered by Dautry and Nyman (DN)\cite{dau,kut,tak}, where $\sigma$ and $\pi^0$ 
meson condensates take the same form as Eq.~(\ref{chiral}). 
The same configuration has been also assumed for non-uniform chiral phase 
in hadron matter by the use of the Nambu-Jona-Lasinio model \cite{sad}.
However, DCDW is by no means the pion condensation but should
be directly considered as particle-hole and particle-antiparticle quark
condensation in the deconfinement phase. }
. This configuration looks to break the translational invariance as well
as rotation symmetry, but the former invariance is recovered by absorbing
an additional constant by a global chiral transformation. 
Accordingly, we define a new quark field $\psi_W$ by the Weinberg 
transformation \cite{wei},
\beq
\psi_W=\exp[i\gamma_5\tau_3 {\bf q\cdot r}/2 ]\psi,
\label{wein}
\eeq
to separate the degrees of freedom of the amplitude and phase of  
DCDW in the Lagrangian. In terms of the new field the effective Lagrangian 
renders
\beq
{\cal L}_{MF}=\bar\psi_W[i\ds{\partial}-M-1/2\gamma_5\tau_3\ds{q}]\psi_W
-G\Delta^2,
\label{effl}
\eeq
where we put $M\equiv -2G\Delta$ and $q^\mu=(0, {\bf q})$, taking the
chiral limit ($m_c=0$). 
The form given in (\ref{effl}) appears to be the same as 
the usual one,  except the axial-vector field 
generated by the wave vector of DCDW; the {\it amplitude} of DCDW 
produces the dynamical quark mass in this case.  
We shall see 
the wave vector ${\bf q}$ is related to the magnetization: the 
{\it phase} of DCDW induces the magnetization.
With this form we can find a 
spatially uniform solution for the quark wave function (see Table~\ref{wein}),
$\psi_W=u_W(p)\exp(i{\bf p\cdot r})$, 
with the eigenvalues,
\beq
E_p^{\pm}=\sqrt{E_{p}^{2}+|{\bf q}|^2/4\pm \sqrt{({\bf
p}\cdot{\bf q})^2+M^{2}|{\bf q}|^2}},~~~E_p=(M^2+|{\bf p}|^2)^{1/2}
\label{energy}
\eeq
for positive-energy (valence) quarks with different spin polarizations 
(c.f. (\ref{eig})). 
\footnote{This feature is very different from refs.\cite{der}, where wave 
function is no more uniform.}  

\begin{table}[h]
\caption{Diagram of the Weinberg transformation.}
\centering
\begin{tabular*}{10cm}{@{\extracolsep{\fill}}lcc}
$\langle\bar\psi \psi\rangle\neq 0$~~~&$\Longleftrightarrow$&~~~
$\langle\bar\psi_W \psi_W\rangle=\Delta(\neq 0)$\\
&&\\
$\langle\bar\psi i\gamma_5\tau_3\psi\rangle\neq 0$~~~&&~~~
$\langle\bar \psi_W i\gamma_5\tau_3\psi_W\rangle=0$\\
&&\\
&& ~~~$ q/2\propto \nabla\theta$~~~(``{\rm axial-vector}'')\\
&&\\
{\rm non-uniform} &&~~~~~~~~~{\rm uniform}
\end{tabular*}
\label{wein}
\end{table}

\subsection{Thermodynamic potential}

The thermodynamic potential is given as
\beqa
\Omega_{\rm total}&=&\gamma\int\frac{d^3p}{(2\pi)^3}\sum_{s=\pm}
\left[(E^s_p-\mu)\theta_s-E^s_p\right]
+M^2/4G\nonumber\\
&\equiv&\Omega_{\rm val}+\Omega_{\rm vac}+M^2/4G .
\label{therm}
\eeqa
where $\theta_\pm=\theta(\mu-E^\pm_p)$, $\mu$ is the chemical potential and 
$\gamma$ the degeneracy factor $\gamma=N_fN_c$. The first term 
$\Omega_{\rm val}$ is the 
contribution by the valence quarks filled up to the chemical potential, 
while the second term $\Omega_{\rm vac}$ is the vacuum 
contribution that is apparently divergent. We shall see both contributions
are {\it indispensable} in our discussion.  
Once $\Omega_{\rm total}$ is properly evaluated, the equations to be solved to 
determine the optimal values of $\Delta$ and $q$ are 
\beq
\frac{\delta\Omega_{\rm total}}{\delta\Delta}
=\frac{\delta\Omega_{\rm total}}{\delta q}=0.
\label{self}
\eeq

Since NJL model is not renormalizable, we need some regularization procedure 
to get a meaningful finite value for the vacuum contribution
$\Omega_{\rm vac}$, which can be recast in the form,
\beq
\Omega_{\rm vac}=i\gamma\int\frac{d^4p}{(2\pi)^4}{\rm trln}S_W,
\eeq
with use of the propagator $S_W=(\ds{p}-M-1/2\tau_3\gamma_5\ds{q})^{-1}$.
There are various kinds of regularization and we must carefully choose
the relevant one to the theoretical framework.
Since the energy spectrum
is no more rotation symmetric, we cannot apply the usual energy or momentum cut-off 
regularization (MCOR)
scheme to regularize $\Omega_{\rm{vac}}$. Moreover, 
the regularization should be, at
least, independent of the order parameters $\Delta$ and $q$. 
Note that 
this demand is essential to discuss the phase transition: 
improper regularizations spoil the consistency of the framework and 
give {\it unphysical} results for the order
parameters $\Delta$ and $q$ 
through Eq.~(\ref{self}). We adopt here the 
proper-time 
regularization (PTR) scheme \cite{sch}, which is one of regularizations
compatible with  Eq.~(\ref{self})
\footnote{The Pauli-Villars reguralization may be another candidate.}
. 
Introducing the proper-time variable $\tau$, we eventually find
\beq
\Omega_{\rm vac}=\frac{\gamma}{8\pi^{3/2}}\int_0^\infty
\frac{d\tau}{\tau^{5/2}}
\int^\infty_{-\infty}\frac{dp_z}{2\pi}\left[
e^{-(\sqrt{p_z^2+M^2}+q/2)^2\tau}
+e^{-(\sqrt{p_z^2+M^2}-q/2)^2\tau}\right]-\Omega_{\rm ref},
\label{j}
\eeq
which is reduced to the standard formula \cite{kle} 
in the limit $q\rightarrow 0$.

The integral with respect to the proper time $\tau$ is still divergent 
due to the $\tau\sim 0$ contribution.
Regularization proceeds by replacing the lower bound of the integration range 
by $1/\Lambda^2$, which corresponds to the momentum cut-off in the 
MCOR scheme.

Now we examine a possible instability of quark matter with respect to
formation of DCDW. 
In the following we first inquire the sign change of the 
curvature of $\Omega_{\rm total}$ at the origin ({\it stiffness} parameter), 
$\beta$.
Expanding $\Omega_{\rm vac}$ with respect to $q$ up to $O(q^2)$, we find
\beq
\Omega_{\rm vac}=\Omega_{\rm vac}^0+\beta_{\rm vac}q^2+O(q^4)
\label{l}
\eeq
where the vacuum stiffness parameter $\beta_{\rm vac}$ is given by 
\beq
\beta_{\rm vac}=\frac{\gamma\Lambda^2}{16\pi^2}J(M^2/\Lambda^2)
\eeq
with a universal function,
$
J(x)=-x{\rm Ei}(-x).
$
The nontrivial term originates from 
a vacuum 
polarization effect in the 
presence of DCDW and  
provides a 
kinetic term ($\propto (\nabla\theta)^2$) for DCDW
. The vacuum stiffness parameter $\beta_{\rm vac}$ can be also written
as 
$\beta_{vac}=\frac{1}{2}f_\pi^2$ \cite{kle} with the pion decay constant
$f_\pi$, and is always positive; 
it gives a 'repulsive' contribution, 
so that the vacuum is stable against 
formation of DCDW. Note that it gives a null contribution 
in case of $M=0$ 
, irrespective of $q$, as it should be. 

For given $\mu, M$ and $q$ we can evaluate the contribution by the Fermi
seas 
$\Omega_{\rm val}$ using 
Eq.~(\ref{energy}), 
but its general formula  is very complicated \cite{tat04}. However, it may be sufficient to 
consider the small $q$ case for our present purpose. 
Then the thermodynamic potential can be expressed as
\beqa
\Omega_{\rm val}&=&\Omega_{\rm val}^0
-\frac{\gamma}{8\pi^2}M^2q^2H(\mu/M)+O(q^4)\nonumber\\
&\equiv& \Omega_{\rm val}^0+\Omega_{\rm val}^{mag}+O(q^4)
\label{xb}
\eeqa 
up to $O(q^2)$, where $H(x)={\rm ln}(x+\sqrt{x^2-1})$ and $\Omega_{\rm val}^0
=\epsilon_{\rm val}^0-\mu\rho_{\rm val}^0$ with 
$\rho^0_{\rm val}=\frac{\gamma}{3\pi^2}(\mu^2-M^2)^{3/2}$ for normal
quark matter.
The valence stiffness parameter then reads
\beq
\beta_{val}=-\frac{\gamma}{8\pi^2}M^2H(\mu/M)
\eeq
Since the function $H(x)$ is always positive and accordingly 
$\beta_{val}\leq 0$,
the magnetic term $\Omega_{\rm val}^{mag}$ always gives a negative energy and 
approaches to zero as $M\rightarrow 0$ (triviality).  

\begin{figure}[h]
\begin{center}
\includegraphics[width=5cm,keepaspectratio]{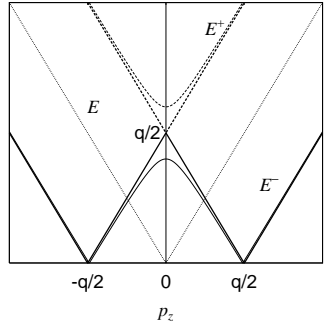}
\end{center}
 \caption{Energy spectra for ${\bf p}_{\perp}=0$. $E^\pm$ with $M=0$ 
(thick solid and dashed lines).
$\tilde E^\pm$ with the 
definite chirality is also shown for comparison (dotted line). 
We can see there is a degeneracy of $E^\pm$ 
at $p_z=0$ for $M=0$, while it is resolved by the mass 
(thin solid and dashed lines).}
\label{level}
\end{figure}

We 
may easily understand why the valence quarks always favor the formation of 
DCDW. 
First, consider the energy spectra for massless quarks (see Fig.~\ref{level}). 

As is 
already discussed, our theory becomes trivial in this case and we find two 
spectra   
\beq
E^\pm_p=\sqrt{p_\perp^2+(|p_z|\pm q/2)^2}, ~~~{\bf p}_\perp=(p_x,p_y, 0),
\label{zeom}
\eeq
which are essentially equivalent to $E^\pm_p=|{\bf p}|$ with definite chirality.

There 
is a level crossing at ${\bf p}={\bf 0}$. Once the mass term is taken into 
account this degeneracy is resolved and the energy splitting arises there. 
Hence it causes an energy gain, if $q=O(2\mu)$; we can 
see that this  
mechanism is very similar to that of SDW by Overhauser \cite{der,ove}.

Using Eqs.~(\ref{therm}), (\ref{l}), (\ref{xb}) we write the thermodynamic 
potential as 
\beq
\Omega_{\rm total}=\Omega_{NJL}+\beta q^2+O(q^4)
\label{xf}
\eeq
with the total stiffness parameter $\beta=\beta_{vac}+\beta_{val}$ and the usual 
NJL expression without DCDW,
$
\Omega_{NJL}=\Omega_{\rm vac}^0(M)+\Omega_{\rm val}^0(M)+M^2/4G.
$
The dynamical quark mass $M$ is 
given by the equation, $\partial\Omega_{\rm total}/\partial M=0$;
At the order of $q^0$ the dynamical quark mass $M^0$ is determined by the equation,
$
\left.\partial\Omega_{NJL}/\partial M\right|_{M^0}=0.
$
Since $M-M^0=O(q^2)$, DCDW onsets 
at a certain density where the total stiffness parameter $\beta$  
becomes negative:  the critical chemical 
potential $\mu^{cr}$ is determined by the equation,
\beq
\beta
=\frac{1}{2}f_\pi^2-\frac{\gamma}{8\pi^2}\left(M^0\right)^2H(\mu^{cr}/M^0)=0.
\label{y}
\eeq
 Note that this is only a {\it sufficient} condition for 
formation of DCDW, and we can {\it never} exclude the 
possibility of the first-order phase transition or metamagnetism \cite{tat00,blo}.    
Actually, we shall see that DCDW occurs as a first-order phase transition.

\subsection{First-order phase transition}

\begin{figure}
\vspace*{-0cm}
\begin{center}
\includegraphics[width=7cm, angle=-90]{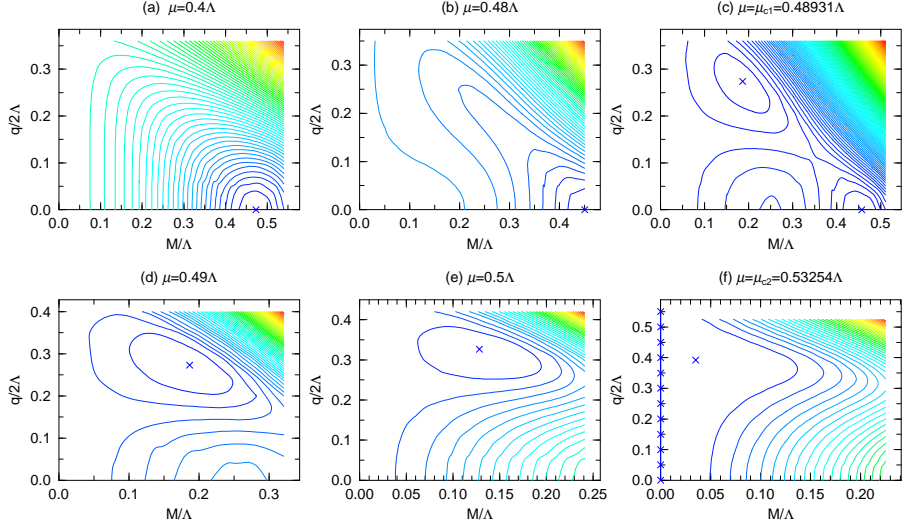}
\end{center}
\caption{Contours of $\Omega_{\rm total}$ at $T=0$ are shown in $M-q$ plane 
as the chemical potential increases, (a) $\rightarrow$ (f).  
The cross in each figure denotes the absolute minimum.}
\label{cp1}
\end{figure}

The values of the order parameters $M$ and $q$ are obtained from the minimum 
of the thermodynamic potential ~(\ref{therm}) for $T=0$.  
Fig.~\ref{cp1} shows the contours of $\Omega_{\rm total}$ in the $M$-$q$ plane  
as the chemical potential increases,   
where the parameters are chosen as $G\Lambda^2=6$ and $\Lambda=850$ MeV, 
to reproduce the constituent quark mass in the vacuum ($\mu=0$)  \cite{kle}.  

The crossed points denote the absolute minima. 
There are two critical chemical potential $\mu=\mu_{c1}, \mu_{c2}$:    
for the lower densities (Fig.~\ref{cp1}(a)-(b)) 
the absolute minimum resides at the point $(M\neq 0, q=0)$ 
indicating the SSB phase. 
At $\mu=\mu_{c1}$ (Fig.~\ref{cp1}(c)) 
the potential has the two absolute minima at $(M\neq 0, q=0)$ and $(M\neq 0, q\neq 0)$, 
showing the first-order transition to the DCDW phase   
which is stable for $\mu_{c1}<\mu<\mu_{c2}$ (Fig.~(\ref{cp1})d-e). 
At $\mu=\mu_{c2}$ (Fig.~\ref{cp1}(f))   
the axis of $M=0$ and a point $(M\neq 0, q\neq 0)$ become minima, 
the system undergoes the first-order transition again to the chiral-symmetric phase.    

\begin{figure}[h]
\begin{minipage}{0.48\textwidth}
\begin{center}
\includegraphics[width=5cm]{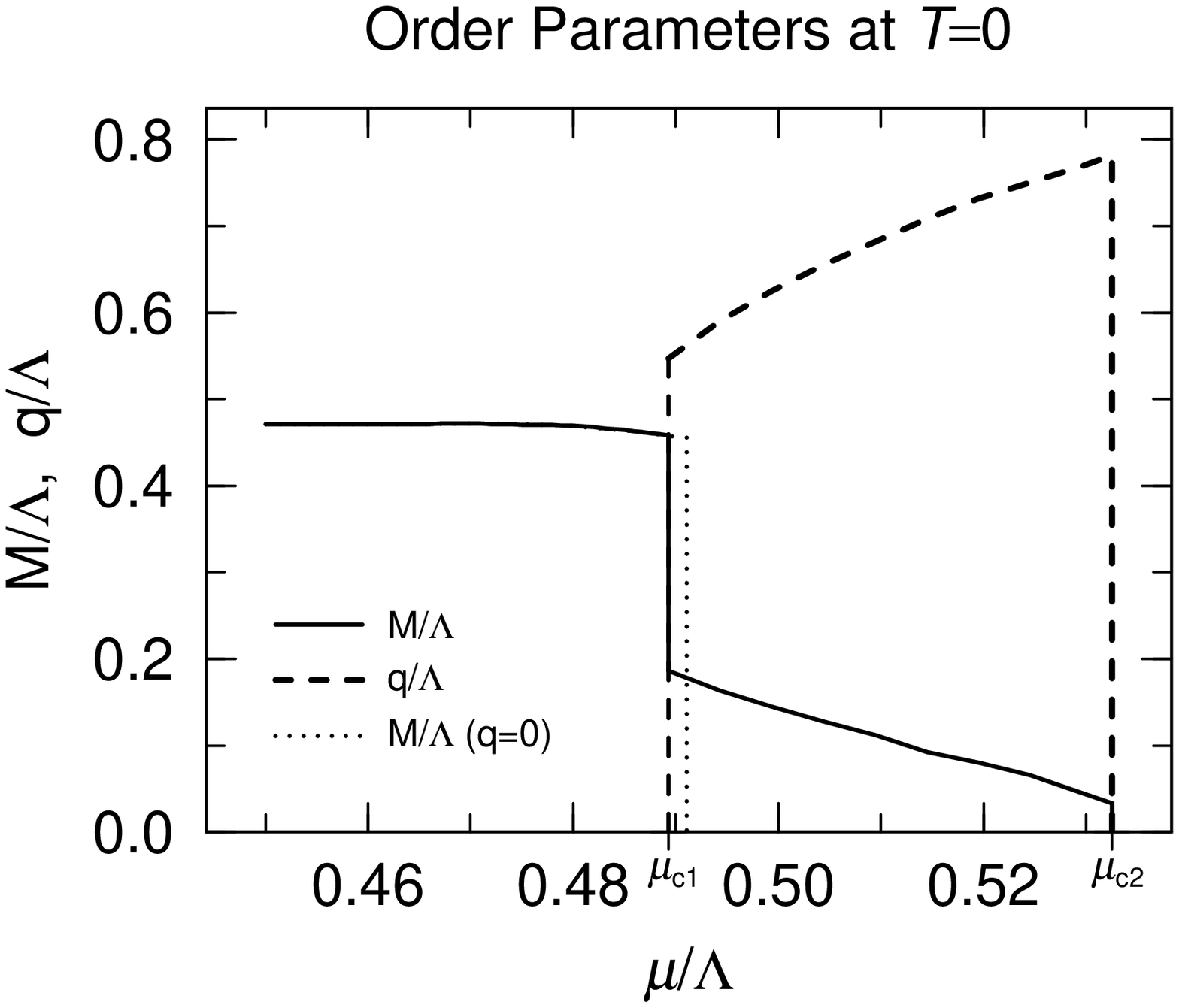}
\end{center}
\caption{Wave number $q$ and the dynamical mass $M$ are plotted 
as functions of the chemical potential at $T=0$. 
Solid (dotted) line for $M$ with (without) the density wave, 
and dashed line for $q$.}
\label{op1}
\end{minipage}
\hspace{\fill}
\begin{minipage}{0.48\textwidth}
\begin{center}
\includegraphics[width=5cm]{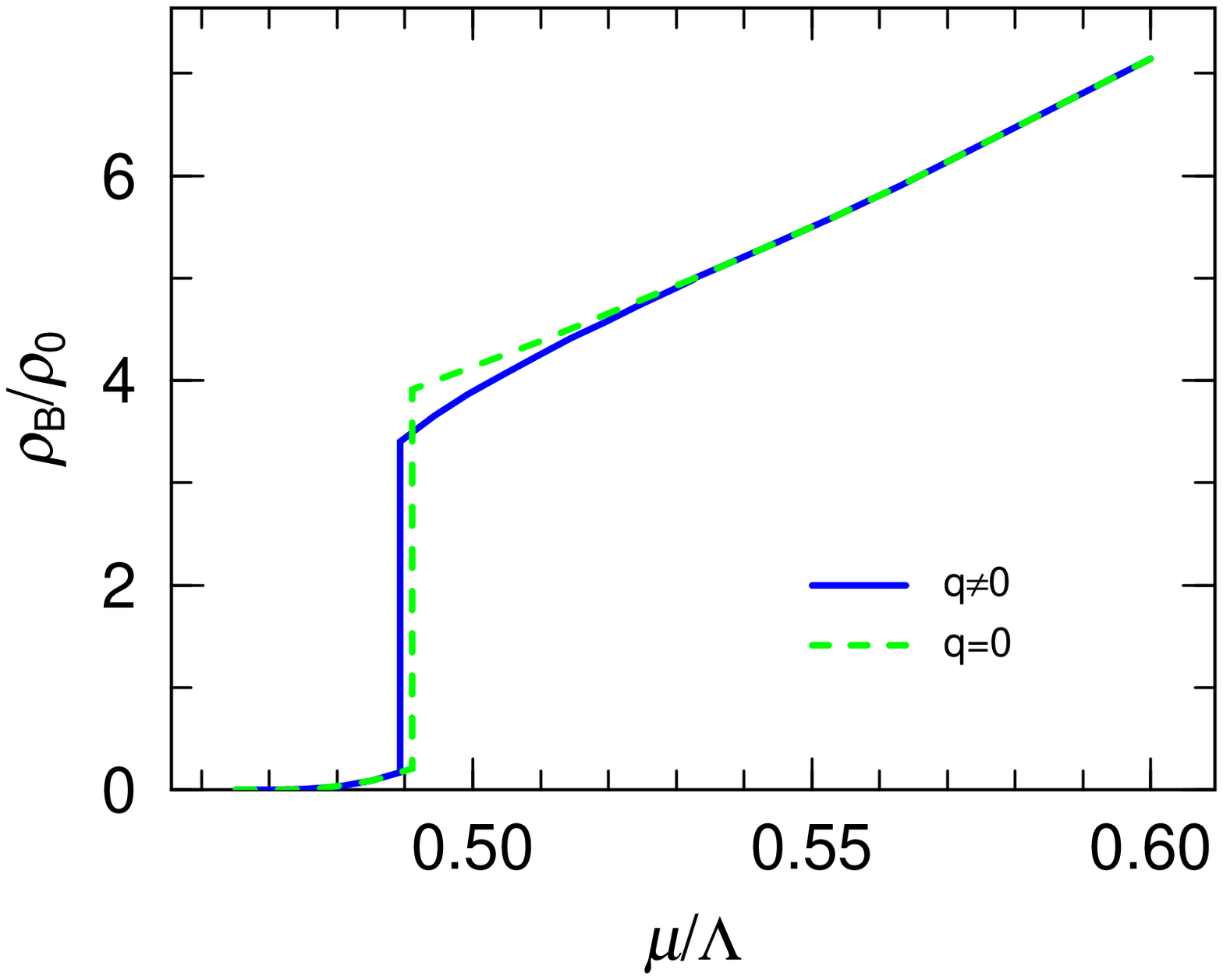}
\end{center}
\vspace*{-0.5cm}
\caption{Baryon number density as a function of $\mu$. 
$\rho_0=0.16 {\rm fm}^{-3}$: the normal nuclear density.}
\label{rhoCDW}
\end{minipage}
\end{figure}

Fig.~\ref{op1} summarizes the behaviors of the order-parameters $M$ and $q$ 
as functions of $\mu$ at $T=0$, 
where that of $M$ without DCDW is also shown for comparison. 
It is found that 
DCDW develops at finite range of $\mu$ ($\mu_{c1}\le\mu\le\mu_{c2}$), where  
the wave number $q$ increases with $\mu$ but its value 
is smaller than twice of the Fermi momentum 
$2k_F$($\simeq 2\mu$ for free quarks) 
since  
the nesting of Fermi surfaces is incomplete in the present 3-D system; 
actually, the ratio becomes $q/k_F=1.17-1.47$ 
for the baryon-number densities
$\rho_b/\rho_0=3.62-5.30$ where DCDW is stable (see Fig.~\ref{op1}).  

\subsection{Correlation functions} 
In this section, 
we consider scalar- and pseudoscalar-correlation functions, $\Pi_{\rm s, sp}(k)$, 
in the massless limit $M\rightarrow 0$,  
and discuss their relation with the mechanism for DCDW. 
In the static limit $k_0\rightarrow 0$, 
the correlation functions have a physical correspondence 
to the static susceptibility for the spin- or charge-density wave \cite{kagoshima}. 
We shall see that these functions have a differential singularity at $k=2k_F$,
reflecting the sharp Fermi surface at $T=0$. 

We explicitly evaluate the effective interactions, $\Gamma_{\rm s, sp}(k)$, in the
pseudo-scalar and scalar channels     
within the random phase approximation \cite{nam, kle}, which  
are related to the correlation functions $\Pi_{\rm s, sp}(k)$, 
i.e.,  $2G\Pi_{\rm s, sp}(k)=\Gamma_{\rm s, sp}(k)\Pi_{\rm s, sp}^0(k)$:  
\begin{eqnarray}
i\Gamma_{\rm s, ps}(k)
&=& \frac{2Gi}{1-2 G\Pi_{\rm s, ps}^0(k)}, 
\end{eqnarray}  
where $\Pi_{\rm s, ps}^0(k)$ are the polarization functions in medium,    
\begin{eqnarray}
\!\!\!\!\!\!\!\!\!\!\Pi_{\rm s}^0(|\bk|)\!\!\!\!&=&\!\!\!\!\Pi_{\rm ps}^0(|\bk|)\nonumber\\ 
\!\!\!\!\!&=&\!\!\!\!\!\frac{N_f N_c }{4\pi^2}(\Lambda^2-2k_F^2)-2N_f N_c i\bk^2I(\bk^2)|_{M\rightarrow0} \nn 
\!\!\!\!\!&+&\!\!\!\!\!\frac{N_f N_c |\bk|}{4\pi^2}\left[\!
\left(k_F\!-\!\frac{|\bk|}{2}\right)\log\left( \frac{2k_F\!+\!|\bk|}{2k_F\!-\!|\bk|} \right) 
\!+\!\frac{|\bk|}{2}\log\left( \frac{2k_F}{|\bk|}\!+\!\frac{|\bk|}{2k_F} \right) 
\!\right]\!\!,\!\! 
\end{eqnarray}  
in the static and chiral limit \cite{tat04}. It is well known that poles
of the effective interaction give the energies of scalar and
pseudo-scalar mesons  \cite{nam, kle}.
Note that the inverse of the effective interaction in the massless limit
also gives   
the coefficient of $M^2$ in the effective potential in the presence of DCDW, 
\begin{eqnarray}
\Omega_{\rm total}=\Omega_{\rm total}|_{M\rightarrow0}+\frac{1}{2}\Gamma_{\rm ps}^{-1}(q)|_{M\rightarrow0} M^2+O(M^4). 
\label{expM}
\end{eqnarray} 
Hence, the critical density and the critical wave vector $q_{\rm crit}$
should be given by the equations,
\beq
\Gamma_{\rm ps}^{-1}(q_{\rm crit})|_{M\rightarrow0}=0,
\partial\Gamma_{\rm ps}^{-1}(q)|_{M\rightarrow0}/\partial q|_{q_{\rm crit}}=0,
\eeq
in the case of the second-order or weakly first-order phase transition. 
\begin{figure}
\begin{center}
\includegraphics[height=5.5cm]{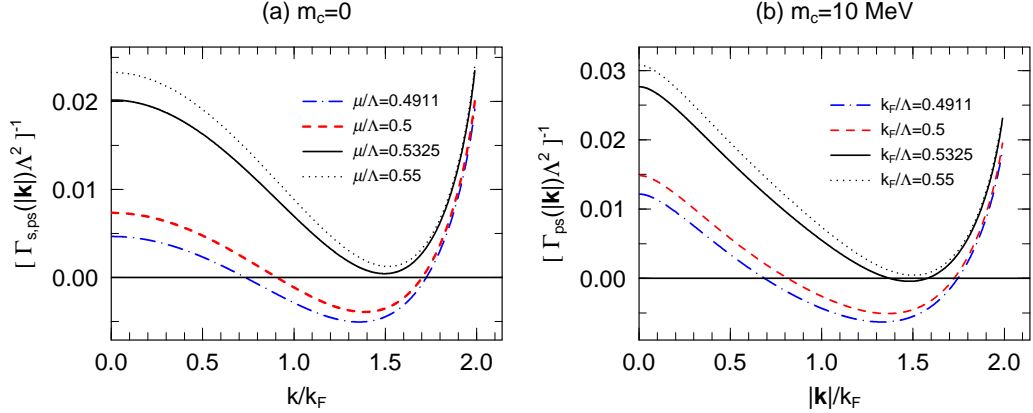}
\end{center}
\caption{Function, $1/\Gamma_{\rm ps}(|\bk|)$, is plotted for various Fermi momenta, 
$k_F/\Lambda=0.4752, 0.5$, and $0.6$. 
The thick (thin) lines correspond to the chiral limit $m_c=0$ ($m_c=5$MeV). 
In the case of $k_F/\Lambda\ge 0.4752$, the mass-gap equation has a
 extremum solution $M=0$ in the absence of DCDW. }
\label{corrfn1}
\end{figure}
Fig.\ref{corrfn1} show the function $1/\Gamma_{\rm ps}(|{\bf
k}|)|_{M=0}$ and 
we can see these conditions are almost satisfied at the terminal density,
$\mu=\mu_{c2}$, due to a tiny jump in the dynamical mass 
(weakly first-order phase transition)
. A numerical calculation in the chiral limit gives 
$\mu_t(=k_F)=0.5320 \Lambda$ and $|\bk_t|=1.498 k_F$, 
which almost coincide with the previous results given in Figs.~\ref{cp1}
and \ref{op1},  
$\mu_{c2}=0.53254 \Lambda$ and $q=1.469 k_F$ 
(where $k_F\equiv\sqrt{\mu_{c2}^2-M^2}$; $M=0.034 \Lambda$).     
On the other hand, the phase
transition is of first order at the onset density, $\mu=\mu_{c1}$: there is a
discontinuous jump in the dynamical mass (the amplitude of DCDW) $M$, so
that the above argument cannot be applied any more. 
Besides, the correlation functions or the effective
interactions  provide a powerful tool to analyze the DCDW phase 
as far as the dynamically generated mass (the amplitude of DCDW) $M$ is
small, as in the present case.
From the behavior of the function $\Gamma_{\rm ps}(|\bk|)^{-1}$ shown in Fig.~{\ref{corrfn1}}, 
it is found that $\Gamma_{\rm ps}(|\bk|)^{-1}$ takes the lowest value
at $|\bk| \sim 1.3-1.5k_F(O(2k_F))$, reflecting
the sharp Fermi surface, 
\footnote{Actually the function $\Gamma_{\rm ps}(|\bk|)^{-1}$ diverges at
$|\bk|=2k_F$ in the one-dimensional case, which means the complete nesting.}
and thus a finite wave number $q$ gives the lower potential energy in
Eq.~(\ref{expM}) than $q=0$ in the density range of DCDW.  These values
are consistent with those in Fig.~\ref{op1}, and we can see again that
DCDW is closely related to the sharpness of the Fermi surface.

It should be noted again that the negative value of the function
$\Gamma_{\rm ps}(|\bk|)^{-1}|$ 
gives a necessary condition for formation of DCDW, but 
the sign change does not necessarily imply the critical condition in the case of first-order phase 
transitions, as in the present case;  
the terminal transition is weakly first-order and we can apply there.
It should be also noted that its minimum point always gives an optimal value of the wave vector in
the presence of DCDW. Thus we can see by 
the use of the correlation functions that 
the particle-hole pairing with finite momentum $q=O(2k_F)$ effectively lowers the free energy in 
comparison with the zero total momentum.

The above argument might also be available 
even for the case of a finite current-quark mass, $m_c\simeq 5$MeV: 
Fig.~{\ref{corrfn1}} shows that the minimum of $\Gamma_{\rm
ps}(|\bk|)^{-1}$ 
has little shift from that in the chiral limit.

\subsection{Magnetic properties}

The mean-value of the spin operator is given by 
\beq
\bar s_z=\frac{1}{2}u^\dagger_W\Sigma_z u_W
=\frac{1}{2}\frac{q/2\pm\beta_p}{E^\pm_p}+{\rm vac},
\label{spin}
\eeq
with $\beta_p=\sqrt{p_z^2+m^2}$, where "vac" means the vacuum contribution.
First note that the integral of $\bar s_z$ over the Fermi seas should be proportional to $q$, and 
the solution with $q\neq 0$ seems to imply FM. However, we can show that PTR 
gives the 
vacuum (the Dirac sea) contribution oppositely to cancel the total mean-value 
of the spin operator, which is consistent with Eq.~(\ref{self}). 
Instead we can see that the magnetization spatially 
oscillates, 
\beq
M_z\equiv \langle \bar q\sigma_{12}q\rangle=\langle\gamma_0\sigma_{12}\rangle \cos({\bf q}\cdot{\bf r}),
\eeq
with 
\beq
\langle\gamma_0\sigma_{12}\rangle=\int_{F^+-F^-}\frac{d^3p}{(2\pi)^3}\frac{2M}
{\sqrt{M^2+p_z^2}},
\eeq
which means a kind of spin density wave \cite{gru}.

\subsection{Phase diagram in the $T-\mu$ plane}
To establish the phase diagram in the  $T-\mu$ plane,  
we derive the thermodynamic potential at finite temperature 
in the Matsubara formalism. 
The partition function for the mean-field Hamiltonian is given by  
\begin{eqnarray}
Z_\beta \!\!&=&
\!\!\!\int {\it D}\bar{\psi} {\it D}\psi 
\exp \int_0^\beta \!\!\!d\tau\!\!\!\int d^3r ~
 \!\!\left\{ \bar{\psi}\left[
i \tilde{\partial} +M \exp\left(i\gamma_5 {\bf q} \cdot {\bf r}\right) 
\!-\!\gamma_0 \mu \right]\psi 
\!-\!\frac{M^2}{4 G} \right\}\nonumber\\ 
&=& 
\prod_{\bk,n,s=\pm} 
\left\{ (i \omega_n+\mu)^2-E_s^2(\bk) \right\}^{N_fN_c}
\times \exp\left\{-\left( \frac{M^2}{4 G}\right)V\beta \right\},
\end{eqnarray}
where $\beta=1/T$, 
$\tilde{\partial}\equiv -\gamma_0 \partial_\tau+i {\bf \gamma}\nabla$ and 
$\omega_n$ the Matsubara frequency. 
Thus the thermodynamic potential $\Omega_\beta$ is obtained,   
\begin{eqnarray}
\Omega_\beta (q,M)\!\!\!
&=&\!\!\!-T \log{Z_\beta(q,M)}/V \nn
\!\!\!&=&\!\!\!-N_fN_c\!\!\int\!\!\frac{d^3{\rm k}}{(2\pi)^3}\!\!\sum_s\!\!\left\{
T\log{\left[ e^{-\beta\left(E_s(\bk)-\mu \right)}\!\!+\!\!1 \right] 
      \left[ e^{-\beta\left(E_s(\bk)+\mu \right)}\!\!+\!\!1 \right]}
\!\!+\!\!E_s(\bk) \right\}\nonumber\\
&&+\frac{M^2}{4 G}\label{tpot}. 
\end{eqnarray}

\begin{figure}[h]
\begin{center}
\includegraphics[width=8cm]{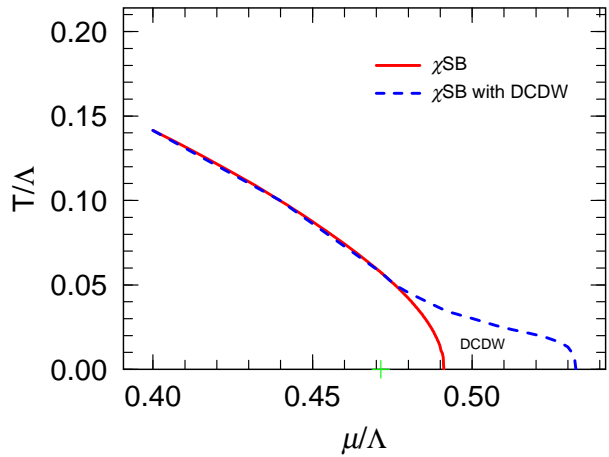}
\end{center}
\caption{Phase diagram obtained 
from the thermodynamic potential ~(\ref{tpot}). 
The solid (dashed) line shows the chiral-restoration boundary  
in the presence (absence) of DCDW. 
The closed area denoted by ``DCDW'' shows the DCDW phase. }
\label{pd1}
\end{figure}

From the absolute minima of the thermodynamic potential (\ref{tpot}), 
it is found that 
the order parameters at $T\neq0$ behave similarly to those at $T=0$ 
as a function of $\mu$,  
while the chemical-potential range of the DCDW at finite temperature, 
$\mu_{c1}(T)\le \mu \le\mu_{c2}(T)$,  
gets smaller as $T$ increases. 
We show the resultant phase diagram in Fig.~\ref{pd1}, 
where the ordinary chiral-transition line is also given. 
Comparing phase diagrams with and without $q$,  
we find that the DCDW phase emerges in the area 
(closed area in Fig.~\ref{pd1})
which lies just outside the boundary of the ordinary chiral transition.
We thus conclude that the DCDW is induced by finite-density contributions, 
and has an effect to expand the chiral-condensed phase ($M\neq 0$) 
toward low temperature and high density region. 

\section{Summary and Concluding remarks}

We have seen some magnetic aspects of quark matter: ferromagnetism at 
high densities and spin density wave at moderate densities within the 
zero-range approximation for the interaction vertex. These look to follow 
the similar development about itinerant electrons: Bloch mechanism at 
low densities and spin density wave at high densities by Overhauser.

By a perturbative calculation 
with the OGE interaction, we have seen ferromagnetism in quark matter at
low densities (\S 2.1)
. It would be worth mentioning that another 
study with higher-order diagrams qualitatively supports it \cite{nie}. 
These studies suggest an opposite 
tendency to the one using the zero-range interaction (\S\S 
2.2,2.3). Note that we can also see the 
same situation for itinerant electrons; the Hartree-Fock calculation based 
on the infinite-range Coulomb interaction favors ferromagnetism at a low 
density region, while the Stoner model, which introduces the zero-range effective 
interaction instead of the Coulomb interaction, gives ferromagnetism at high 
densities. So we must carefully examine the possibility of ferromagnetism in 
quark matter by taking into account the finite-range effect.   
 
We have examined the coexistence of spin polarization 
and color superconductivity by choosing a quark pair with the same polarization.
We have introduced the axial-vector self-energy 
and the quark pair field (the gap function), 
whose forms are derived from the one-gluon-exchange interaction 
by way of the Fierz transformation 
under the zero-range approximation. 
Within the relativistic Hartree-Fock framework 
we have evaluated their magnitudes 
in a self-consistent manner by way of the coupled Schwinger-Dyson equations.

As a result of numerical calculations  
spontaneous spin polarization occurs
at a high density for a finite quark mass 
in the absence of CSC, 
while it never appears for massless quarks as an analytical result.
In the spin-polarized phase the single-particle energies corresponding
to 
spin degrees of freedom, which are degenerate in the non-interacting
system,  
are split by the exchange energy in the axial-vector channel. 
Each Fermi sea of the single-particle energy deforms 
in a different way, 
which causes an asymmetry in the two Fermi seas and then  
induces the axial-vector mean-field in a self-consistent manner.   
In the superconducting phase, however, 
spin polarization is slightly reduced by the pairing effect; 
it is caused by 
competition between reduction of the deformation and 
enhancement of the difference in 
the phase spaces of opposite ``spin'' states 
due to the anisotropic diffuseness in the momentum distribution. 

We have also noted another possibility of the pairing: the quark pair
with opposite polarization to each other. It may lead to a gapless
superconductor, but we need a further study.  

We have seen that dual chiral desnity wave (DCDW) appears at a certain density 
and develops at moderate densities (\S 3). It occurs as a result of the interplay  
between the $\bar qq$ and particle-hole correlations. The phase 
transition is of weakly first order, and the restoration of chiral symmetry is 
delayed compared with the usual scenario. 

For the discussion of DCDW given in \S 4, we have seen the remarkable 
roles of the Fermi sea and the Dirac sea: the former always favors  
DCDW, while the latter works against it. The similar 
situation also appears about the magnetic property of quark matter. The mean 
value of the spin operator over the Fermi seas of valance quarks always gives a
finite value 
in the presence of DCDW, which is a kind of 
ferromagnetism, but the vacuum contribution given by 
the Dirac seas completely cancels it. As a result there is no net spin 
polarization in this case, but we have seen magnetization spatially oscillates 
instead (spin density wave). This is one of the typical examples in which the nonrelativistic 
picture is qualitatively different from the relativistic one by the vacuum 
effect.

It would be interesting to recall that DCDW  
is similar to pion condensation within the $\sigma$ model, 
considered by Dautry and Nyman \cite{dau}, where $\sigma$ and $\pi^0$ 
meson condensates take the same form as Eq.~(\ref{chiral}). So it might
be intriguing to connect pion condensation before deconfinement with DCDW
after it in light of symmetry consideration. Note
that this type of hadron-quark continuity has been also suggested 
in the context
of hadron and quark superconductivities \cite{qhc}.

If ferromagnetism is realized in quark matter, it may give a microscopic
origin of the magnetic field in compact stars; actually we have seen
that it can give a possible explanation for the superstrong magnetic
field observed in magnetars, if they are quarks stars. It would be
challenging to explain other characteristic phenomena in magnetars such as a sudden
braking down observed in a soft gamma-ray repeater SGR 1806-20 or 
SGR 1900+14 \cite{kou}; some
global reconfiguration of of the magnetic has been suggested for these 
phenomena \cite{woo}. 
It would be also ambitious to give a scenario based on magnetic properties of quark 
matter, which can explain the hierarchy of the magnetic field observed 
in three classes of neutron stars, magnetars, radio pulsars and recycled 
millisecond pulsars. Ferromagnetism may give a permanent magnetization 
and there is no field decay, in difference from the dynamo mechanism
caused by the charged current.

The magnetic phases considered here accompany the symmetry breaking,
$SO(3)\rightarrow O(2)$, so that we can expect the Nambu-Goldstone modes
as lowest excitations in the ground state: spin wave in ferromagnetism
and phason in DCDW. It should be interesting to study these modes. 
Such low excitation modes may affect the thermal
evolution of compact stars \cite{tsu}. It would be also interesting to
investigate how the effective interaction by exchanging such excitations
between quarks affects superconductivity.

\section*{Acknowledgments}

This work is partially supported by the Grant-in-Aid for the 21st Century COE
``Center for the Diversity and Universality in Physics '' from 
the Ministry of Education, Culture, Sports, Science and
Technology of Japan. It is also partially supported by the Japanese 
Grant-in-Aid for Scientific
Research Fund of the Ministry of Education, Culture, Sports, Science and
Technology (13640282, 16540246).


\begin{thebibliography}{99}
\bibitem{alf} M. Alford, Ann. Rev. Nucl. Part.Sci., {\bf 51} (2001)
	131.\\
D.H. Rischke, Prog. Part. Nucl. Phys. {\bf 52} (2004) 197.

\bibitem{lat} For recent reviews of lattice simulations, F. Karsh,
	Lect. Notes Phys. {\bf 583} (2002) 209.\\
E. Laermann and O. Philipsen, hep-ph/0303042.

\bibitem{tat05} F. Weber, {\it Pulsars as Astrophysical Laboratories for
	Nuclear and Particle Physics} (IOP Publishing Ltd., Bristol,
	1999).\\
N. Glendenning, {\it Compact Stars} (Springer, New York, 1996).

\bibitem{bai}
D. Bailin and A. Love, Phys. Reports 107(1984) 325.\\
As a review, K. Rajagopal and F. Wilczek, hep-ph/0011333.

\bibitem{BCS} M. Tinkham, {\it Introduction to Superconductivity}
	(McGraw-Hill, New York, 1975).

\bibitem{MAG3} For a review, G. Chanmugam,
		  Annu. Rev. Astron. Astrophys. {\bf 30} (1992) 143.

\bibitem{tho04} P.M. Woods and C.J. Thompson, astro-ph/0406133.

\bibitem{ibr02} A.I. Ibrahim et al., Astrophys. J. {\bf 574} (2002) L51 \\
N. Rea et al., Astrophys. J. {\bf 586} (2003) L65.\\
G.F. Bignami et al., astro-ph/0306189.

\bibitem{mak03} K. Makishima, Prog. Theor. Phys. Suppl. {\bf 151} (2003) 54.

\bibitem{fan01} For recent results, S. Fantoni et al., Phys. Rev. Lett. {\bf 87} (2001)
	181101;
I. Vidana and I. Bombaci, Phys. Rev. {\bf C66} (2002) 045801.

\bibitem{tat00} T. Tatsumi, Phys. Lett. {\bf B489} (2000) 280.

\bibitem{CSC3} M. Alford, K. Rajagopal and F. Wilczek, 
               Nucl. Phys. {\bf B537} (1999) 443; 
               J. Berges and K. Rajagopal, Nucl. Phys. {\bf B538} (1999) 215; 
               T. M. Schwarz, S. P. Klevansky and G. Papp, 
               Phys. Rev. {\bf  C60} (1999) 055205;
               D.Ebert, V.V.Khudyakov, V.Ch.Zhukovsky and K.G.Klimenko, 
               Phys. Rev. {\bf  D65} (2002) 054024.

\bibitem{MagSup1} L.N. Buaevskii et al., Adv. Phys. {\bf 34} (1985) 175.

\bibitem{MagSup2} S.S. Sexena et al., Nature {\bf 406} (2000) 587; 
                  C.Pfleiderer et al., Nature {\bf 412} (2001) 58; 
                  N.I.Karchev et al., Phys. Rev. Lett. {\bf 86} (2001) 846. 

\bibitem{der0}
D.V. Deryagin, D. Yu. Grigoriev and V.A. Rubakov, Int. J. Mod. Phys. {\bf A7}
(1992) 659.

\bibitem{der}
B.-Y. Park, M.Rho, A.Wirzba and I.Zahed, Phys. Rev. {\bf D62} (2000) 034015.\\
R. Rapp, E.Shuryak and I. Zahed, Ohys. Rev. {\bf D63} (2001) 034008.

\bibitem{kagoshima}
 S. Kagoshima, H. Nagasawa, and T. Sambongi, 
{\it One Dimensional Conductors}, 
Springer series in solid-state sciences, Vol. 72
(Springer-Verlag, Berlin, 1988); 
L. P. Gor'kov and G. Gr\"uner, {\it Charge Density waves in Solids}, 
MODERN PROBLEMS IN CONDENSED MATTER SCIENCES. VOL. 25
(AMSTERDAM: North-Holland, 1989).  

\bibitem{peiel1}
R. E. Peierls, {\it Quantum Theory of Solids} 
(Oxford University Press, London, 1955). 

\bibitem{gru1} G. Gr\"uner, Rev. Mod. Phys. {\bf 60} (1988) 4.

\bibitem{ove}
A.W. Overhauser, Phys. Rev. {\bf 128} (1962) 1437.

\bibitem{gru}
G. Gr\"uner, Rev. Mod. Phys. {\bf 66} (1994) 1.

\bibitem{blo} F. Bloch, Z. Phys. {\bf 57} (1929) 545;

\bibitem{yoshi} C. Herring, {\it Exchange Interactions among Itinerant
	Electrons: Magnetism IV} (Academic press, New
	York, 1966)\\
K. Yoshida, {\it Theory of magnetism} (Springer, Berlin, 1998).

\bibitem{bru} K.A. Brueckner and K. Sawada, Phys. Rev. {\bf 112}
(1957) 328.

\bibitem{cep} D.M. Ceperley and B.J. Alder, Phys. Rev. Lett. {\bf 45}
	(1980) 566.

\bibitem{you} D.P. Young et al., Nature {\bf 397} (1999) 412.


\bibitem{itz} e.g.,C. Itzykson and J.-B. Zuber, {\it Quantum Field
Theory} (McGraw-Hill Inc., 1980). 

\bibitem{ber} V.B. Berestetsii, E.M. Lifshitz and L.P. Pitaevsii,\\ 
{\it Relativistic Quantum Theory} (Pergamon Press, 1971).

\bibitem{bay} G. Baym and S.A. Chin, Nucl. Phys. {\bf A262} (1976)
527.

\bibitem{akh} I.A. Akhiezer and S.V. Peletminskii, JETP {\bf 11} (1960) 1316.

\bibitem{tam} e.g.,R. Tamagaki and T. Tatsumi,
Prog. Theor. Phys. Suppl. {\bf 112} (1993) 277.

\bibitem{deg} T. DeGrand et al., Phys. Rev. {\bf D12} (1975)2060.

\bibitem{far} E. Farhi and R.L. Jaffe, Phys. Rev. {\bf D30} (1984) 2379.

\bibitem{nie} A. Niegawa, hep-ph/0404252.

\bibitem{mad} For a review, J. Madsen, astro-ph/9809032.

\bibitem{LeBe} J.I.Kapusta, {\it Finite-Temperature field theory}
               (Cambridge Univ. Press, 1989);  
               M. LeBellac, {\it Thermal Field Theory}
               (Cambridge Univ. Press, Cambridge, Eingland, 1996) 

\bibitem{nak05} E. Nakano, K. Nawa and T. Tatsumi, in preparation.

\bibitem{nak} E. Nakano, T. Maruyama and T. Tatsumi, Phys. Rev. {\bf D68}
	        (2003) 105001; T. Tatsumi, T. Maruyama and E. Nakano,
	Prog. Theor. Phys. Suppl. {\bf 153} (2004) 190; hep-ph/0312351.

\bibitem{scha} T. Schafer, Phys. Rev. {\bf D62} (2000) 094007.\\
M. Buballa, J. Hosek and M. Oertel, Phys. Rev. Lett. {\bf 90} (2003)
	182002.\\
A. Schmitt, Q. Wang and D.H. Rischke, Phys. Rev. {\bf D66} (2002)
	114010;
Phys. Rev. Lett. {\bf 91} (2003) 242301;Phys. Rev. {\bf D69} (2004) 094017.

                  
\bibitem{leg} A. J. Leggett, Rev. Mod. Phys. 47 (1975) 331.

\bibitem{NM3P} R. Tamagaki, Prog. Theor. Phys. {\bf 44} (1970) 905; 
               M. Hoffberg, A.E. Glassgold, R.W. Richardson and M. Ruderman,
               Phys. Rev. Lett. {\bf 24} (1970) 775.

\bibitem{alf3} M. Alford, J.A. Bowers, J.M. Cheyne and G.A. Cowan, hep-ph/0210106.

\bibitem{Ripka} G.Ripka, {\it Quarks Bound by Chiral Fields}
                (Oxford Univ. Press, 1997).

\bibitem{nam}
Y. Nambu and G. Jona-Lasinio, Phys. Rev. {\bf } (1961) 345.

\bibitem{ful} P. Fulde and R.A. Ferrell, Phys. Rev. {\bf 135} (1964)
	550.\\
A.I. Larkin and Y. Ovchinikov, Sov. Phys.-JETP {\bf 20} (1965) 762.

\bibitem{cas} For a review article, R. Casalbuoni and G. Nardulli, Rev. Mod. Phys. {\bf 76}
	(2004) 263.

\bibitem{alf2} M. Alford, J. Berges and K. Rajagopal,
	Phys. Rev. Lett. {\bf 84} (2000) 598.\\
W.V. Liu and F. Wilczeck, Phys. Rev. Lett. {\bf 90} (2003) 047002.\\
I. Shovkovy and H. Huang, Phys. Lett. {\bf B564} (2003) 205.\\
E. Gubankova, W.V. Liu and F. Wilczek, hep-ph/0304016.\\
M.G. Alford, J. Kouvaris and K. Rajagopal, hep-ph/0311286.

\bibitem{tin} 
For a recent review, 
 M. Huang, hep-ph/0409167.

\bibitem{bed} P.F. Bedaque, H. Caldas and R. Rupak,
	Phys. Rev. Lett. {\bf 91} (2003) 247002.\\
H. Caldas, hep-ph/0312275.

\bibitem{sho} I. Shovkovy, M. Hanauske and M. Huang, Phys. Rev. {\bf
	D67} (2003) 103004.\\
S. Reddy and R. Rupak, nucl-th/0405054. 

\bibitem{naw} K. Nawa, E. Nakano and T. Tatsumi, in progress.

\bibitem{kar} N. Karchev, Phys. Rev. {\bf B67}(2003) 054416.

\bibitem{tat04} T. Tatsumi and E. Nakano, hep-ph/0408294; E. Nakano and
	T. Tatsumi, hep-ph/0411350.

\bibitem{sug}
T. Eguchi and H. Sugawara, Phys. Rev. {\bf D10} (1974) 4257.\\
K. Kikkawa, Prog. Theor. Phys. {\bf 56} (1974) 947.

\bibitem{kle}
S.P. Klevansky, Rev. Mod. Phys. {\bf 64} (1992) 649.\\
T. Hatsuda and T. Kunihiro, Phys. Rep. {\bf 247} (1994) 221.

\bibitem{dau}
F. Dautry and E.M. Nyman, Nucl. Phys. {\bf 319} (1979) 323.\\

\bibitem{kut}
M. Kutschera, W. Broniowski and A. Kotlorz, Nucl. Phys. 
{\bf A516} (1990) 566.

\bibitem{tak}
K. Takahashi and T. Tatsumi, Phys. Rev. {\bf C63} (2000) 015205;
Prog. Theor. Phys. {\bf 105} (2001) 437.

\bibitem{sad}
M. Sadzikowski and W. Broniowski, Phys. Lett. {\bf 488} (2000) 63.\\
M. Sadzikowski, Phys. Lett. {\bf 553} (2003) 45.

\bibitem{wei} 
S. Weinberg, {\it The quantum theory of field
	II}(Cambridge, 1996).

\bibitem{sch}
J. Schwinger, Phys. Rev. {\bf 92} (1951) 664.

\bibitem{qhc} 
T. Schaefer and F. Wilczek, Phys. Rev. Lett. {\bf 82} (1999) 3956. 

\bibitem{kou} C. Kouveliotou et al., Nature {\bf 393} (1998) 235; Ap.J
	{\bf 510} (1999) L115.

\bibitem{woo} P.M. Woods et al., astro-ph/0101045.

\bibitem{tsu} S. Tsuruta, Phys. Reports {\bf 292} (1998) 1.\\
D.G. Yakovlev and C.J. Pethick, astro-ph/002143.

\end{thebibliography}
\end{document}